\documentclass[aps,a4paper,superscriptaddress,showpacs,preprintnumbers,amsmath,amssymb]{revtex4}

\usepackage{ulem}

\usepackage{psfrag} \usepackage{graphicx} \usepackage{dcolumn}
\usepackage{color} \usepackage{latexsym,amsfonts} \usepackage{bm}
\usepackage{amssymb}
\usepackage{amsmath}
\usepackage{bbm}
\begin{document}

\title{Electrodisintegration of Deuteron \\ into Dark Matter and Proton
  Close to Threshold}

\author{A. N. Ivanov}\email{ivanov@kph.tuwien.ac.at}
\affiliation{Atominstitut, Technische Universit\"at Wien, Stadionallee
  2, A-1020 Wien, Austria}
\author{R.~H\"ollwieser}\email{roman.hoellwieser@gmail.com}
\affiliation{Atominstitut, Technische Universit\"at Wien, Stadionallee
  2, A-1020 Wien, Austria}\affiliation{Department of Physics,
  Bergische Universit\"at Wuppertal, Gaussstr. 20, D-42119 Wuppertal,
  Germany} \author{N. I. Troitskaya}\email{natroitskaya@yandex.ru}
\affiliation{Atominstitut, Technische Universit\"at Wien, Stadionallee
  2, A-1020 Wien, Austria}
\author{M. Wellenzohn}\email{max.wellenzohn@gmail.com}
\affiliation{Atominstitut, Technische Universit\"at Wien, Stadionallee
  2, A-1020 Wien, Austria} \affiliation{FH Campus Wien, University of
  Applied Sciences, Favoritenstra\ss e 226, 1100 Wien, Austria}
\author{Ya. A. Berdnikov}\email{berdnikov@spbstu.ru}\affiliation{Peter
  the Great St. Petersburg Polytechnic University, Polytechnicheskaya
  29, 195251, Russian Federation}

\date{\today}

\begin{abstract}
We discuss an investigation of the dark matter decay modes of the
neutron, proposed by Fornal and Grinstein (Phys. Rev. Lett.  {\bf
  120}, 191801 (2018)) and Ivanov {\it et al.} ( arXiv:1806.10107
[hep-ph]) for solution of the neutron lifetime anomaly problem,
through the analysis of the electrodisintegration of the deuteron $d$
into dark matter fermions $\chi$ and protons $p$ close to
threshold. We calculate the triple--differential cross section for the
reaction $e^- + d \to \chi + p + e^-$ and propose to search for such a
dark matter channel in coincidence experiments on the
electrodisintegration of the deuteron $e^- + d \to n + p + e^-$ into
neutrons $n$ and protons close to threshold with outgoing electrons,
protons and neutrons in coincidence. A missing of neutron signals
should testify a detection of dark matter fermions.
\end{abstract} 

\maketitle

\section{Introduction}
\label{sec:introduction}

Recently Fornal and Grinstein \cite{Fornal2018, Fornal2019a,
  Fornal2019b, Fornal2020a, Fornal2020b} have proposed a solution to
the neutron lifetime anomaly (NLA) problem, related to a discrepancy
between experimental values of the neutron lifetime measured in bottle
and beam experiments, through a contribution of the neutron dark
matter decay mode $n \to \chi + e^- + e^+$, where $\chi$ is a dark
matter fermion and $(e^- e^+)$ is the electron--positron
pair. However, according to experimental data \cite{Tang2018,Sun2018,
  Klopf2019}, the decay mode $n \to \chi + e^- + e^+$ is
suppressed. So at first glimpse it seems that the decay mode $n \to
\chi + e^- + e^+$ cannot explain the NLA problem. In order to overcome
such a problem we have assumed \cite{Ivanov2018d} that an
unobservability of the decay mode $n \to \chi + e^- + e^+$ may only
mean that the production of the electron--positron pair in such a
decay is below the reaction threshold, i.e. a mass $m_{\chi}$ of dark
matter fermions obeys the constraint $ m_{\chi} > m_n - 2 m_e$, where
$m_n$ and $m_e$ are masses of the neutron and electron (positron),
respectively. Then, we have proposed that the NLA problem can be
explained by the decay mode $n \to \chi + \nu_e + \bar{\nu}_e$, where
$(\nu_e \bar{\nu}_e)$ is a neutrino--antineutrino pair
\cite{Ivanov2018d}. Since neutrino $\nu_e$ and electron $e^-$ belong
to the same doublet in the Standard Electroweak Model (SEM)
\cite{PDG2020, DGH2014} (see also \cite{Ivanov2019b, Ivanov2020a}),
neutrino--antineutrino $(\nu_e \bar{\nu}_e)$ pairs couple to the
neutron--dark matter current with the same strength as
electron--positron $(e^-e^+)$ pairs \cite{Ivanov2018d}. We have
extended this effective quantum field theory by a gauge invariant
quantum field theory model of the neutron- and lepton-dark matter
interactions invariant under the $U'_{Y'}(1) \times U''_{Y''}(1)$
gauge symmetry. In the physical phase the dark matter sectors with
$U'_{Y'}(1)$ and $U''_{Y''}(1)$ symmetries are responsible for the
effective interaction ($n \chi \ell\bar{\ell}$) \cite{Ivanov2018d} and
interference of the dark matter into dynamics of neutron stars
\cite{McKeen2018,Motta2018,Baym2018,Cline2018}, respectively. The dark
matter sector with $U''_{Y''}(1)$ symmetry we have constructed in
analogue with scenario proposed by Cline and Cornell
\cite{Cline2018}. This means that dark matter fermions with mass
$m_{\chi} < m_n$ couple to a very light dark matter spin--1 boson
$Z''$ providing a necessary repulsion between dark matter fermions in
order to give a possibility for neutron stars to reach masses of about
$2 M_{\odot}$ \cite{Demorest2010}, where $M_{\odot}$ is the mass of
the Sun \cite{PDG2020}. The corrected versions of the dark matter
sectors, invariant under the $U'_{Y'}(1) \times U''_{Y''}(1)$ gauge
symmetry is expounded in the Appendix.

In connection with different approaches to the explanation of the NLA,
we have to mention another mechanism proposed by Berezhiani
\cite{Berezhiani2017a, Berezhiani2019a}, which is not related to the
decay of the neutron into a dark matter but based on the $n
\longleftrightarrow n'$ transitions, where $n'$ is a mirror neutron
\cite{Berezhiani2017b, Berezhiani2018}. Since, such a mechanism is far
from being applicable to the analysis of the electrodisintegration of
the deuteron, we will not discuss it in this paper. Another mechanism
for the explanation of the NLA proposed by Berezhiani
\cite{Berezhiani2019b}, assuming the existence of the neutron decays
into dark matter particles, is similar to that by Fornal and Grinstein
\cite{Fornal2018, Fornal2019a, Fornal2019b, Fornal2020a,
  Fornal2020b}. So our approach to the NLA seems also to be a
modification of Berezhiani's mechanism as well as a modifications of
the mechanism by Fornal and Grinstein.

We would like to notice that the existence of the reaction $n \to \chi
+ \nu_e + \bar{\nu}_e$ for the explanation of the NLA problem entails
the existence of the reactions $n + n \to \chi + \chi$, $n + n \to
\chi + \chi + \nu_e + \bar{\nu}_e$ and $\chi + \chi \to n + n$, which
together with the reaction $n \to \chi + \nu_e + \bar{\nu}_e$ can
serve as URCA processes for the neutron star cooling
\cite{Gamow1941,Friman1979,Hansel1995}.

We would like to emphasize that a possibility to explain the NLA
problem by the neutron dark matter decay mode $n \to \chi + \nu_e +
\bar{\nu}_e$ is not innocent and demands to pay the following price.
As has been pointed out in \cite{Ivanov2018d}, the explanation of the
neutron lifetime $\tau_n = 888.0(2.0)\,{\rm s}$ \cite{Fornal2018}
within the Standard Model with the axial coupling constant $g_A =
1.2764$ \cite{Abele2018, Czarnecki2018}, reproducing the neutron
lifetime $\tau_n = 879.6(4)\,{\rm s}$ \cite{Ivanov2013}, is not
possible and demands the account for the contributions of interactions
beyond the Standard Model such as the Fierz interference term $b$
\cite{Fierz1937, Jackson1957, Hardy2020, Severijns2019, Abele2019,
  Young2019, Sun2020, Ivanov2019x} equal to $b = - 1.44 \times
10^{-2}$ \cite{Ivanov2019y}. However, as has been shown in
\cite{Ivanov2019y}, the Fierz interference term $b = - 1.44 \times
10^{-2}$ does not contradict the existing experimental data on the
correlation coefficients and asymmetries of the neutron beta decay. It
is obvious that the decay channel $n \to \chi + \nu_e + \bar{\nu}_e$,
since all decay particles are neutral. Hence, such mechanism of the
NLA can be confirmed experimentally by experimental investigations of
reactions, where an emission of a dark matter fermion $\chi$ is
accompanying with emission of charged standard model particles.

Having assumed that the results of the experimental data
\cite{Tang2018,Sun2018,Klopf2019} can be also interpreted as a
production of electron--positron pairs below reaction threshold,
we may assume that the neutron dark matter decay $n \to \chi + \nu_e +
\bar{\nu}_e$ can be confirmed, for example, in the process of the
electrodisintegration of the deuteron into dark matter fermions and
protons $e^- + d \to \chi + p + e^-$ close to threshold, induced by
the $(n\chi e^-e^+)$ interaction \cite{Ivanov2018d}.

The paper is organized as follows. In section \ref{sec:crosssection}
we calculate the triple--differential cross section for the
electrodisintegration of the deutron $e^- + d \to \chi + p + e^-$ into
dark matter fermions $\chi$ and protons $p$. In section
\ref{sec:conclusion} we discuss the obtained results, make an estimate
of the triple--differential cross section, calculated in section
\ref{sec:crosssection}, and propose an experimental observation of
dark matter fermions in coincidence experiments on the
electrodisintegration of the deuteron $e^- + d \to n + p + e^-$ close
to threshold by detecting outgoing electrons, protons and neutrons in
coincidence. A missing of neutron signals should testify an
observation of dark matter fermions. In the Appendix we extend the
gauge invariant and renormalizable effective quantum field theory of
strong and electroweak low-energy interactions, proposed in
\cite{Ivanov2019b, Ivanov2020a}, by the dark matter sector invariant
under $U'_{Y'}(1) \times U''_{Y''}(1)$ gauge symmetry.

\section{Triple--differential cross section for electrodisintegration
 of deuteron into dark matter fermions and protons $e^- + d \to \chi +
 p + e^-$}
\label{sec:crosssection}

For the solution of the neutron lifetime anomaly problem we have
proposed to use the following effective interaction \cite{Ivanov2018d}
\begin{eqnarray}\label{eq:1}
{\cal L}_{\rm DMBL}(x) =
- \frac{G_F}{\sqrt{2}}\,V_{ud}[\bar{\psi}_{\chi}(x)\gamma_{\mu}(h_V +
  \bar{h}_A\gamma^5)\psi_n(x)] [\bar{\Psi}_e(x)\gamma^{\mu}(1 -
  \gamma^5)\Psi_e(x)],
\end{eqnarray}
where $G_F = 1.1664\times 10^{-11}\,{\rm MeV}^{-2}$ is the Fermi weak
coupling constant, $V_{ud} = 0.97370(14)$ is the
Cabibbo-Kobayashi--Maskawa (CKM) matrix element \cite{PDG2020},
extracted from the $0^+ \to 0^+$ transitions \cite{PDG2020}. The
phenomenological coupling constants $h_V$ and $\bar{h}_A$ define the
strength of the neutron-dark matter $n \to \chi$ transitions. Then,
$\psi_{\chi}(x)$ and $\psi_n(x)$ are the field operators of the dark
matter fermion and neutron, respectively. According to the SEM
\cite{PDG2020} (see also \cite{Ivanov2019b}), the field operator
$\Psi_e(x)$ is the doublet with components $(\psi_{\nu_e}(x),
\psi_e(x))$, where $\psi_{\nu_e}(x)$ and $\psi_e(x)$ are the field
operators of the electron--neutrino (electron--antineutrino) and
electron (positron), respectively. The leptonic current
$\bar{\Psi}_e(x)\gamma^{\mu}(1 - \gamma^5)\Psi_e(x)$ has the $V - A$
structure, since electron--neutrinos are practically left--handed.
The amplitude of the reaction $e^- + d \to \chi + p + e^-$ is defined
by
\begin{eqnarray}\label{eq:2}
M(e^-d\to \chi\,p\,e^-)^{\sigma_{\chi},
  \sigma_p,\sigma'_e,}_{\sigma_e, \lambda_d} &=& -
\frac{G_F}{\sqrt{2}}\,V_{ud}\,\langle p(\vec{k}_p, \sigma_p)\chi
(\vec{k}_{\chi}, \sigma_{\chi})|[\bar{\psi}_{\chi}(0)\gamma_{\mu}(h_V
  + \bar{h}_A\gamma^5)\psi_n(0)]|d(\vec{k}_d,
\lambda_d)\rangle\nonumber\\ &&\times\, [\bar{u}_e(\vec{k}\,'_e,
  \sigma'_e)\gamma^{\mu}(1 - \gamma^5) u_e(\vec{k}_e, \sigma_e)],
\end{eqnarray}
where $\lambda_d = 0, \pm 1$ define the polarization states of the
deuteron, $\bar{u}_e(\vec{k}\,'_e, \sigma'_e)$ and
$u_e(\vec{k}_e, \sigma_e)$ are Dirac wave functions of free electrons
in the final and initial states of the reaction. In the matrix element
of the $d \to \chi + p$ transition $\langle p(\vec{k}_p, \sigma_p)\chi
(\vec{k}_{\chi}, \sigma_{\chi})|$ and $|d(\vec{k}_d,
\lambda_d)\rangle$ are the wave functions of the dark matter fermion
and proton in the final state and the deutron in the initial one. They
are defined by \cite{Ivanov2005}
\begin{eqnarray}\label{eq:3}
\langle p(\vec{k}_p, \sigma_p)\chi (\vec{k}_{\chi}, \sigma_{\chi})| =
\langle 0| a_{\chi}(\vec{k}_{\chi}, \sigma_{\chi}) a_p(\vec{k}_p,
\sigma_p)
\end{eqnarray}
and 
\begin{eqnarray*}
\hspace{-0.3in}|d(\vec{k}_d, \lambda_d = \pm 1)\rangle &=&
\frac{1}{(2\pi)^3}\int \frac{d^3q_p}{\sqrt{2
    E_p(\vec{q}_p)}}\frac{d^3q_n}{\sqrt{2 E_n(\vec{q}_n)}}\,\sqrt{2
  E_d(\vec{q}_p + \vec{q}_n)}\,\delta^{(3)}(\vec{k}_d - \vec{q}_p -
\vec{q}_n)\nonumber\\
\hspace{-0.3in}&&\times\,\Phi_{d}\Big(\frac{\vec{q}_p -
\vec{q}_n}{2}\Big)\,
a^{\dagger}_p(\vec{q}_p, \pm 1/2) a^{\dagger}_n(\vec{q}_n, \pm
1/2)|0\rangle,\nonumber\\
\end{eqnarray*}
\begin{eqnarray}\label{eq:4}
\hspace{-0.3in}|d(\vec{k}_d,\lambda_d = 0)\rangle &=&
\frac{1}{(2\pi)^3}\int \frac{d^3q_p}{\sqrt{2
    E_p(\vec{q}_p)}}\frac{d^3q_n}{\sqrt{2 E_n(\vec{q}_n)}}\,\sqrt{2
  E_d(\vec{q}_p + \vec{q}_n)}\,\delta^{(3)}(\vec{k}_d - \vec{q}_p -
\vec{q}_n)\nonumber\\
\hspace{-0.3in}&&\times\,\Phi_{d}\Big(\frac{\vec{q}_p -
  \vec{q}_n}{2}\Big)\, \frac{1}{\sqrt{2}}\,[a^{\dagger}_p(\vec{q}_p, +
  1/2)) a^{\dagger}_n(\vec{q}_n, - 1/2) + a^{\dagger}_p(\vec{q}_p, -
  1/2)) a^{\dagger}_n(\vec{q}_n, + 1/2)]|0\rangle,
\end{eqnarray}
where $|0\rangle$ is the vacuum wave function,
$a^{\dagger}_j(\vec{p}_j, \sigma_j)$ and $a_j(\vec{p}_j, \sigma_j)$
are operators of creation and annihilation of a fermion $j = n,p,\chi$
with a 3--momentum $\vec{p}_j$ and polarization $\sigma_j = \pm 1/2$
obeying standard relativistic covariant anti--commutation relations
\cite{Ivanov2005}. Then, $\Phi_d(\vec{k}\,)$ is the component of the
wave function of the bound $np$--pair in the ${^3}{\rm S}_1$ state
defined in the momentum representation. It is normalized to unity
\cite{Ivanov2005} (see also \cite{Christlmeier2008}):
\begin{eqnarray}\label{eq:5}
\int
|\Phi_d(\vec{k}\,)|^2 d^3k/(2\pi)^3 = 1.
\end{eqnarray}
We neglect the contribution of the component of the wave function of
the bound $np$--pair in the ${^3}{\rm D}_1$--state
\cite{Machleidt1987,Garson2001,Ivanov2001} (see also
\cite{Gilman2002,Christlmeier2008}), which is not important for the
analysis of the electrodisintegration of the deutron into dark matter
fermions and protons. The wave function of the deuteron
Eq.(\ref{eq:4}) is normalized by \cite{Ivanov2005}:
\begin{eqnarray}\label{eq:6}
\langle
d(\vec{k}\,'_d,\lambda\,'_d|d(\vec{k}_d,\lambda_d)\rangle = (2\pi)^3
2E_d(\vec{k}_d)\,\delta^{(3)}(\vec{k}\,'_d -
\vec{k}_d)\,\delta_{\lambda\,'_d \lambda_d}.
\end{eqnarray}
In the non--relativistic approximation for heavy fermions and in the
laboratory frame, where the deuteron is at rest, the amplitudes of the
electrodisintegration of the deuteron into dark matter fermions and
protons are determined by
\begin{eqnarray}\label{eq:7}
&&M(e^-d\to
  \chi\,p\,e^-)^{\sigma_{\chi},\sigma_p,\sigma'_e}_{\sigma_e,
    \lambda_d = \pm 1}= - \sqrt{4 m_d m_n m_{\chi}}\,G_F
  V_{ud}\,\Phi_d(\vec{k}_p)\,\delta_{\sigma_p, \pm 1/2} \,\Big(h_V
  [\varphi^{\dagger}_{\chi}(\sigma_{\chi})\varphi_n(\pm
    1/2)]\nonumber\\ &&\times\,[\bar{u}'_e (\vec{k}^{\,'}_e,
    \sigma'_e)\gamma^0(1 - \gamma^5) u_e(\vec{k}_e, \sigma_e)] -
  \bar{h}_A [\varphi^{\dagger}_{\chi}(\sigma_{\chi})\vec{\sigma}
    \varphi_n(\pm 1/2)]\cdot [\bar{u}'_e (\vec{k}^{\,'}_e,
    \sigma'_e)\vec{\gamma}\,(1 - \gamma^5) u_e(\vec{k}_e,
    \sigma_e)]\Big),\nonumber\\ &&M(e^-d\to
  \chi\,p\,e^-)^{\sigma_{\chi},\sigma_p,\sigma'_e}_{\sigma_e,
    \lambda_d = 0}= - \sqrt{2 m_d m_n m_{\chi}}\,G_F
  V_{ud}\,\Phi_d(\vec{k}_p)\,\Big\{\delta_{\sigma_p, + 1/2} \,\Big(h_V
  [\varphi^{\dagger}_{\chi}(\sigma_{\chi})\varphi_n(-
    1/2)]\nonumber\\ &&\times\,[\bar{u}'_e (\vec{k}^{\,'}_e,
    \sigma'_e)\gamma^0(1 - \gamma^5) u_e(\vec{k}_e, \sigma_e)] -
  \bar{h}_A [\varphi^{\dagger}_{\chi}(\sigma_{\chi})\vec{\sigma}
    \varphi_n(- 1/2)]\cdot [\bar{u}'_e (\vec{k}^{\,'}_e,
    \sigma'_e)\vec{\gamma}\,(1 - \gamma^5) u_e(\vec{k}_e,
    \sigma_e)]\Big)\nonumber\\ && + \delta_{\sigma_p, - 1/2}
  \,\Big(h_V [\varphi^{\dagger}_{\chi}(\sigma_{\chi})\varphi_n(+
    1/2)]\,[\bar{u}'_e (\vec{k}^{\,'}_e, \sigma'_e)\gamma^0(1 -
    \gamma^5) u_e(\vec{k}_e, \sigma_e)] - \bar{h}_A
        [\varphi^{\dagger}_{\chi}(\sigma_{\chi})\vec{\sigma}
          \varphi_n(+ 1/2)]\nonumber\\ &&\cdot [\bar{u}'_e
          (\vec{k}^{\,'}_e, \sigma'_e)\vec{\gamma}\,(1 - \gamma^5)
          u_e(\vec{k}_e, \sigma_e)]\Big)\Big\},
\end{eqnarray}
where $\varphi_{\chi}(\sigma_{\chi})$ and $\varphi_n(\sigma_n)$ are
the Pauli wave functions of the dark matter fermion and neutron,
respectively, $\vec{\sigma}$ are $2\times 2$ Pauli matrices, $m_d =
m_n + m_p + \varepsilon_d$ is the deuteron mass, $m_n$ and $m_p$ are
masses of the neutron and proton, and $\varepsilon_d = -
2.224575(9)\,{\rm MeV}$ is the deuteron binding energy
\cite{Leun1982}, and $m_{\chi}$ is the dark matter fermion mass. The
differential cross section for the reaction $e^- + d \to \chi + p +
e^-$, averaged over polarizations of the incoming electron and
deuteron and summed over polarizations of the fermions in the final
state, is equal to
\begin{eqnarray}\label{eq:8}
 d^9\sigma(E'_e,E_e,\vec{k}^{\,'}_e, \vec{k}_e,
 \vec{k}_{\chi},\vec{k}_p) &=& (1 +
 3 g^2_A)\,\frac{G^2_F|V_{ud}|^2}{8\pi^5}\,\frac{\zeta^{(\rm
     dm)}}{\beta_e}\,\Big(1 + a^{(\rm dm)}\,\frac{\vec{k}^{\,'}_e\cdot
   \vec{k}_e}{E'_e E_e}\Big)\,|\Phi_d(\vec{k}_p)|^2\,\delta(E'_e + E_p
 + E_{\chi} - m_d - E_e)\nonumber\\ &&\times\,
 \delta^{(3)}(\vec{k}_{\chi} + \vec{k}_p + \vec{k}^{\,'}_e -
 \vec{k}_e) \,d^3k_{\chi} d^3k_p d^3k'_e,
\end{eqnarray}
where $\beta_e = k_e/E_e$ is the incoming electron velocity and $g_A =
1.27641(56)$ is the axial coupling constant
\cite{Abele2018,Czarnecki2018} introduced in \cite{Ivanov2018d} for
convenuence.  The correlation coefficients $\zeta^{(\rm dm)}$ and
$a^{(\rm dm)}$ are defined by \cite{Ivanov2018d}
\begin{eqnarray}\label{eq:9}
\zeta^{(\rm dm)} = \frac{1}{1 + 3 g^2_A}\,(|h_V|^2 +
3|\bar{h}_A|^2) = \frac{0.018}{(m_n -
  m_{\chi})^5}\;,\; a^{(\rm dm)} = \frac{|h_V|^2 - |\bar{h}_A|^2
}{|h_V|^2 + 3|\bar{h}_A|^2},
\end{eqnarray}
where $m_n - m_{\chi}$ is measured in MeV. In the non--relativistic
approximation for the dark matter fermion and proton and in the
center--of--mass frame of the $\chi p$--pair we transcribe
Eq.(\ref{eq:8}) into the form
\begin{eqnarray}\label{eq:10}
 d^9\sigma(E'_e,E_e,\vec{k}^{\,'}_e, \vec{k}_e, \vec{p}, \vec{k}\,)
 &=& (1 +
 3\lambda^2)\,\frac{G^2_F|V_{ud}|^2}{8\pi^5}\,\frac{\zeta^{(\rm
     dm)}}{\beta_e}\,\Big(1 + a^{(\rm dm)}\,\frac{\vec{k}^{\,'}_e\cdot
   \vec{k}_e}{E'_e E_e}\Big)\,\Big|\Phi_d\Big(\vec{k} + \frac{m_p}{m_p
   + m_{\chi}}\,\vec{p}\,\Big)\Big|^2\nonumber\\ &&\times
 \,\delta\Big(\frac{p^2}{2M} + \frac{k^2}{2\mu} - ({\cal E}_0 + E_e -
 E'_e)\Big)\,\delta^{(3)}(\vec{p} + \vec{k}^{\,'}_e - \vec{k}_e)\,
 d^3p d^3k d^3k'_e,
\end{eqnarray}
where ${\cal E}_0 = m_n - m_{\chi} + \varepsilon_d$, $\vec{p}$ and
$\vec{k}$ are the total and relative 3--momenta of the $\chi p$--pair,
related to the 3--momenta of the dark matter fermion $\vec{k}_{\chi}$
and the proton $\vec{k}_p$ as follows
\begin{eqnarray}\label{eq:11}
 \vec{k}_ {\chi} = - \vec{k} + \frac{m_{\chi}}{m_p +
   m_{\chi}}\,\vec{p}\quad,\quad \vec{k}_ p = \vec{k} + \frac{m_p}{m_p
   + m_{\chi}}\,\vec{p}.
\end{eqnarray}
Then, $M = m_p + m_{\chi}$ and $\mu = m_p m_{\chi}/(m_p + m_{\chi})$
are the total and reduced masses of the $\chi p$--pair. Having
integrated over $\vec{p}$ we arrive at the expression
\begin{eqnarray}\label{eq:12} 
d^6\sigma(E'_e,E_e,\vec{k}^{\,'}_e, \vec{k}_e, \vec{k}\,) &=&
\frac{\mu}{2\pi^2}\,\frac{\zeta^{(\rm dm)}}{\tau_n f_n\beta_e}\,\Big(1
+ a^{(\rm dm)}\,\frac{\vec{k}^{\,'}_e\cdot \vec{k}_e}{E'_e
  E_e}\Big)\,\Big|\Phi_d\Big(\vec{k} + \frac{m_p}{m_p +
  m_{\chi}}\,\vec{q}\,\Big)\Big|^2\nonumber\\ &&\times
\,\delta\Big(k^2 - 2\mu ({\cal E}_0 + E_e - E'_e) + \frac{\mu}{M}\,q^2
\Big)\,d^3k d^3k'_e,
\end{eqnarray}
where $\vec{q} = \vec{k}_e - \vec{k}^{\,'}_e $ \cite{Arenhovel1982}
and $q^2 = \vec{q}^{\,2}$, and we have used the definition of the
neutron lifetime $1/\tau_n = (1 + 3 g^2_A)\,G^2_F|V_{ud}|^2f_n/2\pi^3$
\cite{Ivanov2013}. Following Arenh\"ovel \cite{Arenhovel1982} and
Arenh\"ovel {\it et al.} \cite{Arenhovel2002} we define the
triple--differential cross section in the center--of--mass frame of
the $\chi p$--pair and in the laboratory frame of incoming and
outgoing electrons
\begin{eqnarray}\label{eq:13} 
\hspace{-0.3in}&&\frac{d^5\sigma(E'_e,E_e,\vec{k}^{\,'}_e, \vec{k}_e,
  \vec{n}\,)}{dE'_e d\Omega_e d\Omega_{\vec{n}}} =
\frac{\mu}{4\pi^2}\,\frac{\zeta^{(\rm dm)}}{\tau_n f_n\beta_e}\,k'_e
E'_e \,\Big(1 + a^{(\rm dm)}\,\frac{\vec{k}^{\,'}_e\cdot
  \vec{k}_e}{E'_e E_e}\Big)\,\Theta\Big(m_p({\cal E}_0 + E_e - E'_e) -
\frac{1}{4}\,q^2\Big)\nonumber\\
\hspace{-0.3in}&&\times\,\sqrt{2\mu ({\cal E}_0 + E_e - E'_e) -
  \frac{\mu}{M}\,q^2}\,\Big|\Phi_d\Big(\vec{n}\sqrt{2\mu ({\cal E}_0 +
  E_e - E'_e) - \frac{\mu}{M}\,q^2}\, + \frac{m_p}{m_p +
  m_{\chi}}\,\vec{q}\,\Big)\Big|^2\,,
\end{eqnarray}
where $\Theta(z)$ is the Heaviside function, $\vec{n} = \vec{k}/k$,
$d\Omega_e = 2\pi\,\sin\theta_e d\theta_e$ and $d\Omega_{\vec{n}} =
\sin\theta d\theta d\phi$ with the standard definition of the
kinematics of the electrodisintegration of the deuteron
\cite{Arenhovel1982,Arenhovel2002} (see Fig.\,1 of
Refs.\cite{Arenhovel1982,Arenhovel2002}), where $\vec{k}^{\,'}_e\cdot
\vec{k}_e = k'_e k_e \cos\theta_e$ and $\vec{n}\cdot \vec{q} =
q\,\cos\theta$. Then, it is convenient to transcribe Eq.(\ref{eq:13})
into the form
\begin{eqnarray}\label{eq:14} 
\hspace{-0.3in}\frac{d^5\sigma(E'_e,E_e,\vec{k}^{\,'}_e, \vec{k}_e,
  \vec{n}\,)}{dE'_e d\Omega_e d\Omega_{\vec{n}}} &=&
\frac{m_p}{8\pi^2}\,\frac{\zeta^{(\rm dm)}}{\tau_n f_n\beta_e}\,k'_e
E'_e \,\Big(1 + a^{(\rm dm)}\,\frac{\vec{k}^{\,'}_e\cdot
  \vec{k}_e}{E'_e E_e}\Big)\,\Theta\Big(m_p({\cal E}_0 +
    E_e - E'_e) - \frac{1}{4}\,q^2\Big)\nonumber\\
\hspace{-0.3in}&&\times\,\sqrt{m_p({\cal E}_0 + E_e - E'_e) -
  \frac{1}{4}\,q^2}\,\Big|\Phi_d\Big(\vec{n}\, \sqrt{m_p({\cal
      E}_0 + E_e - E'_e) - \frac{1}{4}\,q^2}\, +
\frac{1}{2}\,\vec{q}\,\Big)\Big|^2\,.
\end{eqnarray}
For the wave function of the deuteron $\Phi_d(\vec{\ell}\,)$ we may
use the expression
\begin{eqnarray}\label{eq:15} 
\hspace{-0.3in}\Phi_d(\vec{\ell}\,) = \sqrt{\frac{8 \pi}{\displaystyle
    \sum^n_{i = 1}\sum^n_{j = 1}\frac{C_i C_j}{m_i m_j (m_i +
      m_j)^3}}}\,\sum^n_{i = 1} \frac{C_i}{m^2_i + \ell^2}\quad,\quad
\int \frac{d^3\ell}{(2\pi)^3}\,|\Phi_d(\vec{\ell}\,)|^2 = 1
\end{eqnarray}
 with parameters $C_i$ and
$m_i$ taken from the paper by Machleidt {\it et al.}
\cite{Machleidt1987}. Also we may follow Gilman and Gross
\cite{Gilman2002} and describe the wave function
$\Phi_d(\vec{\ell}\,)$ by the expression
\begin{eqnarray}\label{eq:16} 
\hspace{-0.3in}\Phi_d(\vec{\ell}\,) = \frac{\sqrt{8\pi \sqrt{- m_N
      \varepsilon_d}}}{\displaystyle ( - m_N \varepsilon_d +
  \ell^2)\Big(1 + \frac{\ell^2}{p^2_0}\Big)}\,\Big(1 -
\frac{m_N\varepsilon_d}{p^2_0}\Big)^{3/2}\quad, \quad \int
\frac{d^3\ell}{(2\pi)^3}\,|\Phi_d(\vec{\ell}\,)|^2 = 1,
\end{eqnarray}
where $ - m_N\varepsilon_d = 0.940 \times 2.224 \times 10^{-3}\,{\rm
  GeV^2} = 2.09\times 10^{-3}\,{\rm GeV^2}$ and $p^2_0 = 0.15\,{\rm
  GeV^2}$ \cite{Gilman2002}. The squared 3--momenta $\ell^2$ and $q^2$
are defined by
\begin{eqnarray}\label{eq:17}
\ell^2 &=& m_p ({\cal E}_0 + E_e - E'_e) + q\sqrt{m_p({\cal E}_0 + E_e
  - E'_e) - \frac{1}{4}\,q^2}\,\cos\theta,\nonumber\\ q^2 &=&
(\vec{k}_e - \vec{k}^{\,'}_e)^2 = (k_e - k'_e)^2 + 4k'_e k_e
\sin^2\frac{\theta_e}{2}.
\end{eqnarray}
Using the definition of the correlation coefficient $\zeta^{(\rm dm)}$
(see Eq.(\ref{eq:9})) we may rewrite the triple--differential cross
section Eq.(\ref{eq:16}) as follows
\begin{eqnarray}\label{eq:18} 
&&\frac{d^5\sigma(E'_e,E_e,\vec{k}^{\,'}_e, \vec{k}_e, \vec{n}\,)}{dE'_e
  d\Omega_e d\Omega_{\vec{n}}} = \sigma_0 \, \frac{k'_e
  E'_e}{16\pi^2\beta_e}\,\Big(1 + a^{(\rm
  dm)}\,\frac{\vec{k}^{\,'}_e\cdot \vec{k}_e}{E'_e
  E_e}\Big)\,\Theta\Big(m_p({\cal E}_0 + E_e - E'_e) -
\frac{1}{4}\,q^2\Big)\nonumber\\ &&\times\, \sqrt{m_p({\cal E}_0 + E_e -
  E'_e) - \frac{1}{4}\,q^2}\,\Bigg|\Phi_d\Bigg(\sqrt{m_p ({\cal E}_0 +
  E_e - E'_e) + q\sqrt{m_p({\cal E}_0 + E_e - E'_e) -
    \frac{1}{4}\,q^2}\,\cos\theta}\,\Bigg)\Bigg|^2,
\end{eqnarray}
where all momenta and energies and the dimension of the wave function
of the deuteron are measured in ${\rm MeV}$. Then, the scale parameter
$\sigma_0$ is equal to
\begin{eqnarray}\label{eq:19} 
\sigma_0 = 2m_p \frac{\zeta^{(\rm dm)}}{\tau_n f_n} =
6.4\,\Big(\frac{0.12}{m_n - m_{\chi}}\Big)^5\,\frac{\rm fb}{\rm MeV}.
\end{eqnarray}
The triple--differential cross section for the reaction $e^- + d \to
\chi + p + e^-$, given by Eq.(\ref{eq:18}), can be used for the
analysis of the experimental data on searches for dark matter fermions
in coincidence experiments
\cite{Fabian1979,Tamae1987,Neumann-Cosel2002}. For $(m_n - m_{\chi})
\simeq 0.023\,{\rm MeV}$ (see the Appendix and a discussion below
Eq.(\ref{eq:A.23})) the scale parameter $\sigma_0$ increases by four
orders of magnitude $\sigma_0 \simeq 24.7\,{\rm pb/MeV}$.

\section{Discussion}
\label{sec:conclusion}

We have analyzed the electrodisintegration of the deuteron into dark
matter fermions and protons $e^- + d \to \chi + p + e^-$ close to
threshold. Such a disintegration is induced by the electron--neutron
inelastic scattering $e^- + n \to \chi + e^-$ with energies of
incoming electrons larger than the deuteron binding energy
$|\varepsilon_d| = 2.224575(9)\,{\rm MeV}$. The strength of the
reaction $e^- + n \to \chi + e^-$ is caused by the strength of the
neutron dark matter decay mode $n \to \chi + \nu_e + \bar{\nu}_e$,
which has been proposed in \cite{Ivanov2018d} for an explanation of
the neutron lifetime anomaly in case of an unobservability (see
\cite{Tang2018,Sun2018}) of the dark matter decay mode $n \to \chi +
e^- + e^+$ \cite{Fornal2018}, where the production of the
electron--positron pair can be below the reaction threshold
\cite{Ivanov2018d}. Following such an assumption that the production
of the electron--positron pair can be below the reaction threshold for
a confirmation of an existence of the dark matter decay mode $n \to
\chi + e^- + e^+$ we have proposed in \cite{Ivanov2018d} to analyze
the low--energy electron--neutron inelastic scattering $e^- + n \to
\chi + e^-$, which can be in principle distinguished above the
background of the low--energy electron--neutron elastic scattering
$e^- + n \to n + e^-$.

The effective interaction Eq.(\ref{eq:1}) is supported by the
effective quantum field theory model with gauge $SU_L(2) \times U_Y(1)
\times U'_R(1) \times U''_L(1)$ symmetry, where the SM and dark matter
sectors are described by the effective low-energy Lagrangian ${\cal
  L}_{\rm L\sigma M \& SET \& DM' \& DM''}$ (see the Appendix). The SM
part of this effective field theory, determined by the effective
low-energy Lagrangian ${\cal L}_{\rm L\sigma M \& SET}$ invariant
under $SU(2)_L \times U_Y(1)$ gauge symmetry, is gauge invariant and
renormalizable \cite{Ivanov2019b, Ivanov2020a}. This has been
demonstrated in \cite{Ivanov2019b, Ivanov2020a} by examples of the
calculation of the radiative corrections of order $O(\alpha E_e/m_N)$
to the neutron lifetime and correlation coefficients of the neutron
beta decay \cite{Ivanov2021}.  As has been shown in
\cite{Ivanov2018d} (see also the Appendix) such a quantum field theory
model allows i) to derive the effective interaction Eq.(\ref{eq:1}) in
the tree--approximation for the dark matter spin--1 boson $Z'$
exchanges with $h_V = h_A$ (see Eq.(\ref{eq:A.23}) in the Appendix),
and ii) following the scenario by Cline and Cornell \cite{Cline2018}
to show that dynamics of dark matter fermions with mass $m_{\chi} <
m_n \sim 1\,{\rm GeV}$ and a light dark matter spin--1 boson $Z''$,
responsible for a repulsion between dark matter fermions, does not
prevent neutron stars to reach masses of about $2 M_{\odot}$. It has
been also noticed \cite{Ivanov2018d} that the processes $n \to \chi +
\nu_e + \bar{\nu}_e$, $n + n \to \chi + \chi$, $n + n \to \chi + \chi
+ \nu_e + \bar{\nu}_e$ and $\chi + \chi \to n + n$, allowed in such a
quantum field theory model, can serve as URCA processes for the
neutron star cooling \cite{Gamow1941,Friman1979,Hansel1995}. The
effective quantum field theory, described by the Lagrangian ${\cal
  L}_{\rm L\sigma M \& SET \& DM' \& DM''}$ (see the Appendix) is
fully low-energy one. The application of this effective theory to the
analysis of the searches of dark matter in the LHC experiments is not
a straightforward and demands a special consideration, which we are
planing to carry out in our forthcoming publication.

However, in order to have more processes with particles of the
Standard Model in the initial and final states allowing to search dark
matter in terrestrial laboratories we have turned to the analysis of
the dark matter decay mode $n \to \chi + e^- + e^+$ through the
electrodisintegration of the deuteron $e^- + d \to \chi + p + e^-$
induced by the interaction $(n\chi\,e^- e^+)$ \cite{Ivanov2018d}. We
have calculated the triple--differential cross section for the
reaction $e^- + d \to \chi + p + e^-$ (see Eq.(\ref{eq:18})), which
can be used for the analyze of traces of dark matter in coincidence
experiments on the electrodisintegration of the deuteron $e^- + d \to
n + p + e^-$ close to threshold
\cite{Fabian1979,Tamae1987,Neumann-Cosel2002}. An important property
of this cross section is its independence of the azimuthal angle
$\phi$ between the scattering and reaction planes (see Fig.\,1 of
Ref.\cite{Arenhovel2002}). Using the experimental conditions of
Ref.\cite{Neumann-Cosel2002}: $E_e = 50\,{\rm MeV}$, $E_x = E_e - E'_e
= 8\,{\rm MeV}$, $\theta_e = 40^0$, $\theta = 0$ and the wave function
of the deuteron Eq.(\ref{eq:16}) we predict the triple--differential
cross section for the reaction $e^- + d \to \chi + p + e^-$ equal to
\begin{eqnarray}\label{eq:20} 
16\pi^2\frac{d^5\sigma(E'_e,E_e,\vec{k}^{\,'}_e, \vec{k}_e,
  \vec{n}\,)}{dE'_e d\Omega_e d\Omega_{\vec{n}}}\Big|_{E_e, E_x =
  8\,{\rm MeV}, \theta_e = 40^0, \theta = 0} = \Big(\frac{0.12}{m_n
  - m_{\chi}}\Big)^5\, \left\{\begin{array}{r@{}l} ~9\;, &~ E_e =
50\,{\rm MeV}\\ 21\;, &~ E_e = 85\,{\rm MeV}
\end{array}\right.\frac{\rm fb}{\rm MeV},
\end{eqnarray}
where we have set $a^{(\rm dm)} = 0$ \cite{Ivanov2018d}. Following
\cite{Tamae1987} we define the ratio $R(\theta)$ of the
triple--differential cross section at fixed $E_e, E_x, \theta_e$ and
$0 \le \theta \le 180^0$ to the triple--differential cross section at
fixed $E_e, E_x, \theta_e$ and $\theta = 0^0$. We get
\begin{eqnarray}\label{eq:21} 
R(\theta) = \frac{\Big|\Phi_d\Big(\sqrt{m_p ({\cal E}_0
    + E_e - E'_e) + q\sqrt{m_p({\cal E}_0 + E_e - E'_e) -
      \frac{1}{4}\,q^2}\,\cos\theta}\,\Big)\Big|^2}{
  \Big|\Phi_d\Big(\sqrt{m_p ({\cal E}_0 + E_e - E'_e) +
    q\sqrt{m_p({\cal E}_0 + E_e - E'_e) -
      \frac{1}{4}\,q^2}\,}\Big)\Big|^2}.
\end{eqnarray}
\begin{figure}
\includegraphics[height=0.15\textheight]{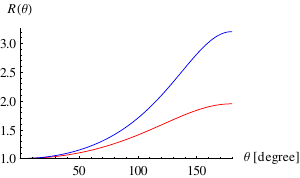}
  \caption{The ratio $R(\theta)$ of the triple--differential cross
    sections for the electrodisintegration of the deuteron $e^- + d
    \to \chi + p + e^-$ for the energies $E_e = 50\,{\rm MeV}$ (red
    curve) and $E_e = 85\,{\rm MeV}$ (blue curve), $E_x = 8\,{\rm
      MeV}$, $E'_e = E_e - E_x$ and $\theta_e = 40^0$
    \cite{Neumann-Cosel2002}.}
\label{fig:fig1} 
\end{figure}
In Fig.\,\ref{fig:fig1} we plot the ratio $R(\theta)$ for the
experimental conditions of \cite{Neumann-Cosel2002} and $0 \le \theta
\le 180^0$.  For $E_e = 50\,{\rm MeV}$ and $E_e = 85\,{\rm MeV}$ the
3--momentum transferred $q$ is equal to $q = 0.16\,{\rm fm^{-1}}$ and
$q = 0.28\,{\rm fm^{-1}}$, respectively.

Of course, the value of the triple--differential cross section
Eq.(\ref{eq:18}) is sufficiently small. Its strong dependence on the
mass difference $m_n - m_{\chi}$ looks rather promising but cannot
guaranty a real enhancement of the cross section.  Nevertheless, we
believe that an observation of the electrodisintegration of the
deuteron into dark matter fermions and protons close to threshold can
be performed in coincidence experiments on the electrodisintegration
of the deutron. Indeed, in \cite{Tamae1987,Neumann-Cosel2002} the
electrodisintegration of the deuteron $e^- + d \to n + p + e^-$ has
been investigated close to threshold detecting outgoing electrons and
protons from the $np$--pairs in coincidence.

We propose to detect dark matter fermions from the
electrodisintegration of the deuteron $e^- + d \to \chi + p + e^-$
above the background $e^- + d \to n + p + e^-$ by detecting outgoing
electrons, protons and neutrons in coincidence.  Since the kinetic
energies of the center--of--mass and relative motion of the $n
p$--pairs (or $\chi p$--pairs) are equal to $(T_{(n/\chi) + p} =
q^2/2M = 0.28\,{\rm MeV},T_{(n/\chi) p} = k^2/2\mu = 5.61\,{\rm MeV})$
and $(T_{(n/\chi) + p} = q^2/2M = 0.83\,{\rm MeV}, T_{(n/\chi) p} =
k^2/2\mu = 5.06\,{\rm MeV})$ for $E_e = 50\,{\rm MeV}$ and $E_e =
85\,{\rm MeV}$, respectively, $E_x = 8\,{\rm MeV}$ and $\theta_e =
40^0$, neutrons should be detected with kinetic energies $T_{(n/\chi)
  p} < 6\,{\rm MeV}$ in the direction practically opposite to the
direction of protons. A missing of neutron signals at simultaneously
detected signals of protons and outgoing electrons should testify an
observation of dark matter fermions in the final state of the
electrodisintegration of the deuteron close to threshold.  The
electron--energy and angular distribution of these events with a
missing of neutron signals should be compared with the distribution
given by Eq.(\ref{eq:18}).

Finally we would like notice that it would be very interesting to
understand an influence of the reactions $n \to \chi + \nu_e +
\bar{\nu}_e$, $e^- + n \to \chi + e^-$, $e^- + d \to \chi + p + e^-$,
$\nu_e(\bar{\nu}_e) + n \to \chi + \nu_e(\bar{\nu}_e)$ and
$\nu_e(\bar{\nu}_e) + d \to \chi + p + \nu_e(\bar{\nu}_e)$, which are
caused by the same interaction Eq.(\ref{eq:1}), on a formation of dark
matter in the Universe during the evolution of the Universe
\cite{HernandezMonteagudo2015,Ejiri2018,Karananas2018}.

\section{Acknowledgements}

We are grateful to Hartmut Abele for fruitful discussions stimulating
the work under this paper. The work of A. N. Ivanov was supported by
the Austrian ``Fonds zur F\"orderung der Wissenschaftlichen
Forschung'' (FWF) under contracts P31702-N27 and P26636-N20 and
``Deutsche F\"orderungsgemeinschaft'' (DFG) AB 128/5-2.  The work of
R. H\"ollwieser was supported by the Deutsche Forschungsgemeinschaft
in the SFB/TR 55.  The work of M. Wellenzohn was supported by the MA
23.

\newpage

\section*{Appendix A: Effective quantum field theory of low-energy
  strong, electroweak and dark matter interactions for the description
  of neutron-dark-matter decays and related processes}
\renewcommand{\theequation}{A-\arabic{equation}}
\setcounter{equation}{0}

\subsection*{\bf Renormalizable and gauge invariant effective quantum
  field theory of strong and electroweak low-energy interactions of
  pions, nucleons, electrons and neutrinos}

The problem of the neutron decays into dark matter is entirely the
low-energy one. In \cite{Ivanov2018d} we have proposed a quantum field
theoretic model of low-energy electroweak and dark matter interactions
for the proton and neutron coupled to the electron and neutrino and
dark matter particles. Unfortunately, the part of this quantum field
theoretic model for the nucleon and leptons cannot be treated as a
hadronized version of the Standard Model (SM) \cite{PDG2020,
  DGH2014}. This problem has been overcome in \cite{Ivanov2019b,
  Ivanov2020a}, where we have proposed the effective quantum field
theory of strong and electroweak low-energy interactions for pions,
proton and neutron coupled to the electron and neutrino that we have
called as the L$\sigma$M$\&$SET.  In this theory strong low-energy
pion-nucleon interactions are described by the linear $\sigma$-model
(L$\sigma$M) with chiral $SU(2) \times SU(2)$ symmetry
\cite{GellMann1960, Lee1972, Nowak1996}. For the description of
electroweak hadron-hadron, hadron-lepton and lepton-lepton
interactions for the electron-lepton family we have used the Standard
Electroweak Theory (SET) with gauge $SU_L(2) \times U_Y(1)$ symmetry
\cite{Weinberg1967, Weinberg1971}. Such an effective quantum field
theory can be treated as a hadronized version of the Standard Model
(SM) \cite{PDG2020, DGH2014} (see \cite{Ivanov2019b, Ivanov2020a}). In
the tree-approximation and in the limit of the infinite mass
$m_{\sigma} \to \infty$ of the $\sigma$-meson the L$\sigma$M
reproduces all results of the current algebra in the form of effective
chiral Lagrangians of pion-nucleon interactions with non-linear
realization of chiral $SU(2)\times SU(2)$ symmetry and different
parametrizations of the pion-field \cite{Weinberg1967a, Weinberg1968,
  Weinberg1979}.  For the exponential parametrization of the
pion-field the Lagrangian ${\cal L}_{\rm L \sigma M}\big|_{m_{\sigma}
  \to \infty}$ of the L$\sigma$M, taken at $m_{\sigma} \to \infty$,
reduces to the Lagrangian ${\cal L}_{\rm HB \chi PT}$ of the chiral
quantum field theory with the structure of low-energy interactions
agreeing well with Gasser-Leutwyler's chiral perturbation theory
(ChPT) or the heavy baryon chiral perturbation theory (HB$\chi$PT)
\cite{Gasser1984, Gasser1987, Gasser1988, Bernard1992, Bernard1995,
  Ecker1995, Ecker1996, Bijnens1996, Bernard1997, Fettes1998,
  Gasser2000, Scherer2002, Scherer2003, Myhrer2005, Bernard2007,
  Bernard2008, Scherer2010, Scherer2011} with chiral $SU(2)\times
SU(2)$ symmetry(see, for example, Ecker \cite{Ecker1995}). Such an
equivalence of the L$\sigma$M with the Gasser-Leutwyler's ChPT has been
also proved in \cite{Bissegger2007} in the leading logarithmic
approximation (see also \cite{Ivanov2020a}).

As has been shown in \cite{Ivanov2019b, Ivanov2020a} the effective
quantum field theory L$\sigma$M$\&$SET is gauge invariant and
renormalizable theory, allowing, for example, a quantitative analysis
of next-to-leading order corrections in the large nucleon mass $m_N$
expansion $O(\alpha E_e/m_N)$ to radiative corrections of order
$O(\alpha/\pi)$, calculated to leading order in the large nucleon
mass expansion by Sirlin \cite{Sirlin1967, Sirlin1978}, Shann
\cite{Shann1971} and Ivanov {\it et al.} \cite{Ivanov2017,
  Ivanov2019a, Ivanov2021}. In this Appendix we extend the effective
quantum field theory L$\sigma$M$\&$SET by the dark matter sectors as
it has been done in \cite{Ivanov2018d}. As a result we obtain a
renormalizable gauge invariant quantum field theoretic model allowing
to take into account the contributions of neutron- and lepton-dark
matter interactions at low energies.

In the symmetric phase the Lagrangian of the effective quantum field
theory L$\sigma$M$\&$SET takes the form \cite{Ivanov2019b}:
\begin{eqnarray}\label{eq:A.1}
\hspace{-0.15in} {\cal L}_{\rm L \sigma M \& SET} &=&
\bar{\Psi}_{NL}i\gamma^{\mu}\Big(\partial_{\mu} +
i\,g\,\frac{1}{2}\,\vec{\tau}_L \cdot \vec{W}_{\mu} +
i\,g'\,\frac{1}{2}\,B_{\mu} \Big)\Psi_{NL} +
\bar{\psi}_{pR}i\gamma^{\mu}\big(\partial_{\mu} +
i\,g'\,B_{\mu})\psi_{pR} +
\bar{\psi}_{nR}i\gamma^{\mu}\partial_{\mu}\psi_{nR}\nonumber\\
\hspace{-0.15in} &-& \sqrt{2}\,g_{\pi
  N}\big(\bar{\Psi}_{NL}\Phi\,\psi_{pR} +
\bar{\psi}_{pR}\Phi^{\dagger}\Psi_{NL}\big) - \sqrt{2}\, g_{\pi
  N}\big(\bar{\Psi}_{NL}\Phi^c\,\psi_{nR} + \bar{\psi}_{nR}\Phi^{c
  \dagger}\Psi_{NL}\big)\nonumber\\
\hspace{-0.15in} &+&  \Big(\partial_{\mu}\Phi^{\dagger} -
i\,g\,\frac{1}{2}\,\Phi^{\dagger}\,\vec{\tau}_L\cdot \vec{W}_{\mu} +
i\,g'\,\frac{1}{2}\,\Phi^{\dagger} B_{\mu}
\Big)\Big(\partial^{\mu}\Phi + i\,g\,\frac{1}{2}\,\vec{\tau}_L\cdot
\vec{W}_{\mu}\Phi - i\,g'\,\frac{1}{2}\,B_{\mu}\Phi \Big) \nonumber\\
\hspace{-0.15in} &+& \mu^2\, \Phi^{\dagger}\Phi -
\frac{1}{2}\,\gamma\, \big(\Phi^{\dagger}\Phi\big)^2 -
\frac{1}{4}\,\vec{W}_{\mu\nu}\cdot \vec{W}^{\mu\nu} -
\frac{1}{4}\,B_{\mu\nu} B^{\mu\nu} + \bar{\Psi}_{\ell
  L}i\gamma^{\mu}\Big(\partial_{\mu} +
i\,g\,\frac{1}{2}\,\vec{\tau}\cdot \vec{W}_{\mu} -
i\,g'\,\frac{1}{2}\,B_{\mu}\Big) \Psi_{\ell
  L}\nonumber\\ \hspace{-0.15in}&+& \bar{\psi}_{eR} i \gamma^{\mu}
\big(\partial_{\mu} - i\,g'\,B_{\mu}\big) \psi_{eR} -
\sqrt{2}\,g_e(\bar{\Psi}_{\ell L}\psi_{eR}\phi +
\phi^{\dagger}\bar{\psi}_{eR}\Psi_{\ell L}) +
\Big(\partial_{\mu}\phi^{\dagger} -
i\,g\,\frac{1}{2}\,\phi^{\dagger}\,\vec{\tau}\cdot \vec{W}_{\mu} -
i\,g'\,\frac{1}{2}\,\phi^{\dagger} B_{\mu}\Big)\nonumber\\
\hspace{-0.15in} &\times&\Big(\partial^{\mu}\phi +
i\,g\,\frac{1}{2}\,\vec{\tau}\cdot \vec{W}^{\,\mu}\phi +
i\,g'\,\frac{1}{2}\,B^{\,\mu}\phi\Big) + \tilde{\mu}^2\,
\phi^{\dagger}\phi - \tilde{\lambda}\,(\phi^{\dagger}\phi)^2,
\end{eqnarray}
where the hadron, lepton and Higgs-boson field operators are defined by
\cite{Ivanov2019b}
\begin{eqnarray}\label{eq:A.2}
\hspace{-0.3in}\Psi_{NL} &=& P_L \psi_N =
P_L\left(\begin{array}{c}\psi_p \\ \psi_n
\end{array}\right)\;,\; \psi_{pR} = P_R\psi_p\; , \; \psi_{nR} = P_R\psi_n,
\nonumber\\ 
\hspace{-0.3in} \Phi &=& \frac{1}{\sqrt{2}}\,\left(\begin{array}{c}
  \sigma + i \pi^3\\ i(\pi^1 + i \pi^2)
\end{array}\right) = \frac{1}{\sqrt{2}}\,\left(\begin{array}{c} \sigma
  + i \pi^0\\ i\,\sqrt{2}\,\pi^-
\end{array}\right),\nonumber\\ 
\hspace{-0.3in} \Phi^c &=& - i\tau_{2L}\Phi^* = \frac{1}{\sqrt{2}}
\,\left(\begin{array}{c} i(\pi^1 - i\pi^2) \\ \sigma - i \pi^3
\end{array}\right) = \frac{1}{\sqrt{2}}\, \left(\begin{array}{c} i\,
  \sqrt{2}\,\pi^+ \\ \sigma - i \pi^0
\end{array}\right),\nonumber\\
\hspace{-0.3in} \Psi_{\ell L} &=& P_L\left(\begin{array}{c} \psi_{\nu_e}
  \\ \psi_e
\end{array}\right)\;,\; \psi_{eR} = P_R\psi_e \;,\;
 \phi  = \left(\begin{array}{c}\phi^+
    \\ \phi^0
\end{array}\right)
\end{eqnarray}
with $P_{L,R} = (1 \mp \gamma^5)/2$ are the left-right projection
operators with the properties $P^2_{L,R}= P_{L,R}$ and $P_LP_R =
P_RP_L = 0$. In Eq.(\ref{eq:A.2}) $\psi_p, \psi_n, \sigma, \pi^{\pm},
\pi^0$ are the field operators of the proton $(p)$, neutron $(n)$,
$\sigma$-meson $(\sigma)$ and pions $(\pi^{\pm}, \pi^0)$,
respectively. Then, $\psi_{\nu_e}, \psi_e, \phi^+, \phi^0$ are the
field operators of the electron neutrino $(\nu_e)$, electron $(e^-)$
and Higgs-bosons $(\phi^+, \phi^0)$, respectively. For the electroweak
gauge boson operators we use the standard notations $\vec{W}_{\mu}$
and $B_{\mu}$ \cite{PDG2020, DGH2014}.  The operators of the field
strength tensors $\vec{W}_{\mu \nu}$ and $B_{\mu \nu}$ of the
electroweak gauge boson fields are given by
\begin{eqnarray}\label{eq:A.3}
\vec{W}_{\mu\nu} &=& \partial_{\mu}\vec{W}_{\nu} -
\partial_{\nu}\vec{W}_{\mu} - g\,\vec{W}_{\mu}\times
\vec{W}_{\nu},\nonumber\\ B_{\mu\nu}&=& \partial_{\mu}B_{\nu} -
\partial_{\nu}B_{\mu}
\end{eqnarray}
The field operators Eq.(\ref{eq:A.1}) have the following properties
under the $SU(2)_L \times U(1)_Y$ infinitesimal transformations
\begin{eqnarray}\label{eq:A.4}
&&\Psi_{NL} \stackrel{\vec{\alpha}_L\,,\, \alpha_Y}\longrightarrow
  \Psi'_{NL} = \Big(1 + i\,\frac{1}{2}\,\vec{\tau}_L\cdot
  \vec{\alpha}_L +
  i\,\frac{1}{2}\,Y\,\alpha_Y\Big)\Psi_{NL},\nonumber\\ &&\psi_{pR}
  \stackrel{\vec{\alpha}_L \,,\, \alpha_Y}\longrightarrow \psi'_{pR} =
  \Big(1 + i\,\frac{1}{2}\,Y\,\alpha_Y\Big)\psi_{pR} \;,\;\psi_{nR}
  \stackrel{\vec{\alpha}_L \,,\,\alpha_Y}\longrightarrow \psi'_{nR} =
  \Big(1 + i\,\frac{1}{2}\,Y\,\alpha_Y\Big)
  \psi_{nR},\nonumber\\ &&\Phi \stackrel{\vec{\alpha}_L
    \,,\,\alpha_Y}\longrightarrow \Phi' = \Big(1 +
  i\,\frac{1}{2}\,\vec{\tau}_L\cdot \vec{\alpha}_L +
  i\,\frac{1}{2}\,Y\,\alpha_Y\Big)\,\Phi,\nonumber\\ &&\Phi^c
  \stackrel{\vec{\alpha}_L \,,\,\alpha_Y}\longrightarrow {\Phi^{c}}' =
  \Big(1 + i\,\frac{1}{2}\,\vec{\tau}_L\cdot \vec{\alpha}_L +
  i\,\frac{1}{2}\,Y\,\alpha_Y\Big)\,\Phi^c,\nonumber\\ && \Psi_{\ell
    L} \stackrel{\vec{\alpha}_L \,,\, \alpha_Y}\longrightarrow
  \Psi'_{\ell L} = \Big(1 + i\,\frac{1}{2}\,\vec{\tau}\cdot
  \vec{\alpha}_L + i\,\frac{1}{2}\,Y\,\alpha_Y\Big)\Psi_{\ell
    L},\nonumber\\ &&\psi_{eR} \stackrel{\vec{\alpha}_L
    \,,\,\alpha_Y}\longrightarrow \psi'_{eR} = \Big(1 +
  i\,\frac{1}{2}\,Y\,\alpha_Y\Big)\psi_{eR},\nonumber\\ &&\phi
  \stackrel{\vec{\alpha}_L \,,\,\alpha_Y}\longrightarrow \phi' =
  \Big(1 + i\,\frac{1}{2}\,\vec{\tau}_L\cdot \vec{\alpha}_L +
  i\,\frac{1}{2}\,Y\,\alpha_Y\Big)\,\phi,\nonumber\\ && \vec{W}_{\mu}
  \stackrel{\vec{\alpha}_L\,,\,\alpha_Y}\longrightarrow \vec{W}'_{\mu}
  = \vec{W}_{\mu} + \vec{W}_{\mu} \times \vec{\alpha}_L -
  \frac{1}{g}\,\partial_{\mu}\vec{\alpha}_L,\nonumber\\ &&B_{\mu}
  \stackrel{\vec{\alpha}_L\,,\,\alpha_Y}\longrightarrow B'_{\mu} =
  B_{\mu} -
  \frac{1}{g'}\,\partial_{\mu}\alpha_Y,\nonumber\\ &&\vec{W}_{\mu\nu}
  \stackrel{\vec{\alpha}_L \,,\, \alpha_Y}\longrightarrow
  \vec{W}\,'_{\mu\nu} = \vec{W}_{\mu\nu} + \vec{W}_{\mu\nu} \times
  \vec{\alpha}_L, \nonumber\\ &&B_{\mu\nu} \stackrel{\vec{\alpha}_L
    \,,\, \alpha_Y}\longrightarrow B'_{\mu\nu} = B_{\mu\nu}
\end{eqnarray}
where $\frac{1}{2}\,\vec{\tau}_L$ (or $\vec{I}_L =
\frac{1}{2}\,\vec{\tau}_L$) and $Y$ are the operators of the {\it weak
  isospin} and {\it weak hypercharge}, respectively, $\vec{\alpha}_L$
and $\alpha_Y$ are infinitesimal parameters of the $SU_L(2)$ and
$U_Y(1)$ gauge group transformations, respectively. The operators of
the third component $I_{3L}$ of the {\it weak isospin} $\vec{I}_L$ and
the {\it weak hypercharge} $Y$ are related by $Q = I_{3L} + Y/2$
\cite{Weinberg1967,Weinberg1971} (see also \cite{PDG2020, DGH2014}),
where $Q$ is the operator of electric charge, measured in the proton
electric charge $e$. The eigenvalues of the third component of the
{\it weak isospin} and {\it weak hypercharge} are $((I_{3L})_{pL},
Y_{pL}) =(+1/2, +1)$, $((I_{3L})_{nL}, Y_{nL}) =(-1/2, +1)$,
$((I_{3L})_{pR},Y_{pR}) = (0, +2)$, $((I_{3L})_{nR},Y_{nR}) = (0, 0)$,
$((I_{3L})_{\sigma + i\pi^0},Y_{\Phi}) =(+1/2, -1)$,
$((I_{3L})_{\pi^-},Y_{\Phi}) =(-1/2, -1)$,
$((I_{3L})_{\pi^+},Y_{\Phi^c}) =(+1/2, + 1)$, and $((I_{3L})_{\sigma -
  i\pi^0},Y_{\Phi^c}) =(-1/2, + 1$, respectively.

At $g = g' = 0$ the effective low-energy Lagrangian ${\cal L}_{\rm
  L\sigma M \& SET}$ reduces to the Lagrangian ${\cal L}_{\rm L\sigma
  M}$ of the linear $\sigma$-model \cite{Ivanov2019b}
\begin{eqnarray}\label{eq:A.5}
\hspace{-0.3in}{\cal L}_{\rm L \sigma M} &=&
\bar{\Psi}_{NL}i\gamma^{\mu}\partial_{\mu}\Psi_{NL} +
\bar{\psi}_{pR}i\gamma^{\mu}\partial_{\mu}\psi_{pR} +
\bar{\psi}_{nR}i\gamma^{\mu}\partial_{\mu}\psi_{nR} - \sqrt{2}\,g_{\pi
  N}\big(\bar{\Psi}_{NL}\Phi\,\psi_{pR} +
\bar{\psi}_{pR}\Phi^{\dagger}\Psi_{NL}\big)\nonumber\\
\hspace{-0.3in} &-& \sqrt{2}\,g_{\pi
  N}\big(\bar{\Psi}_{NL}\Phi^c\,\psi_{nR} + \bar{\psi}_{nR}\Phi^{c
  \dagger}\Psi_{NL}\big) +
\partial_{\mu}\Phi^{\dagger}\partial^{\mu}\Phi + \mu^2\,
\Phi^{\dagger}\Phi - \frac{1}{2} \gamma \big(\Phi^{\dagger}\Phi\big)^2
\end{eqnarray}
invariant under chiral $SU(2) \times SU(2)$ transformations
\cite{Ivanov2019b}. It can be rewritten in the standard form
\cite{GellMann1960, Lee1972, Nowak1996} (see \cite{Ivanov2019b}):
\begin{eqnarray}\label{eq:A.6}
\hspace{-0.15in}{\cal L}_{\rm L\sigma M} =
\bar{\psi}_N\big(i\gamma^{\mu}\partial_{\mu} - g_{\pi N}(\mathbbm{1}
\sigma + i\gamma^5 \vec{\tau}\cdot \vec{\pi}\,)\big) \psi_N +
\frac{1}{2}\,\big(\partial_{\mu}\sigma\partial^{\mu}\sigma +
\partial_{\mu}\vec{\pi}\cdot \partial^{\mu}\vec{\pi}\,\big) +
\frac{1}{2}\,\mu^2\,\big(\sigma^2 + \vec{\pi}^{\,2}\big) -
\frac{1}{4}\,\gamma\,\big(\sigma^2 + \vec{\pi}^{\,2}\big)^2,
\end{eqnarray}
where $\mathbbm{1}$ and $\vec{\tau}$ are isospin $2 \times 2$ unit and
Pauli-like matrices of, respectively. Renormalizability and gauge
invariance of the effective quantum field theory, described by the
Lagrangian ${\cal L}_{\rm L\sigma M\& SET}$ in the spontaneously
broken or physical phases of $SU_L(2)\times U_Y(1)$ and chiral
$SU(2)\times SU(2)$ symmetries has been demonstrated in
\cite{Ivanov2019b, Ivanov2020a} by example of the calculation of i)
the amplitude of the neutron beta decay in the one-hadronic loop
approximation, ii) the radiative corrections of order $O(\alpha
E_e/m_N)$ to the neutron beta decay in the one-loop approximation, and
iii) the radiative corrections of order $O(\alpha E_e/m_N)$ to the
neutron beta decay in the two-loop approximation, where $\alpha$,
$E_e$ and $m_N$ are the fine-structure constant \cite{PDG2020}, the
electron energy and the nucleon mass, respectively.

\subsection*{\bf Dark matter sector of the effective quantum field
  theory of strong, electroweak and dark matter low-energy
  interactions for the description of neutron-dark-matter decays and
  related processes}

Now we are able to add the dark matter sector. According to
\cite{Ivanov2018d}, such a sector contains a dark matter fermion
$\chi$, a dark matter spin--1 gauge boson $C_{\mu}$ and a complex
scalar boson $\varphi$. The Lagrangian of the dark matter sector we
define as follows \cite{Ivanov2018d}
\begin{eqnarray}\label{eq:A.7}
\hspace{-0.3in}&&{\cal L}_{\rm DM'} = \bar{\psi}_{\chi
  R}i\gamma^{\mu}(\partial_{\mu} + i e_{\chi}C_{\mu})\psi_{\chi R} -
\frac{1}{4}\,C_{\mu\nu}C^{\mu\nu} + (\partial_{\mu} -
ie_{\chi}C_{\mu})\varphi^*(\partial_{\mu} + ie_{\chi}C_{\mu})\varphi +
\kappa^2|\varphi|^2 - \gamma |\varphi|^4
\nonumber\\ \hspace{-0.3in}&&+ \bar{\psi}_{\chi
  L}i\gamma^{\mu}\partial_{\mu}\psi_{\chi L} - \sqrt{2}\,
f_{\chi}\Big(\bar{\psi}_{\chi R}\psi_{\chi L} \varphi +
\bar{\psi}_{\chi L}\psi_{\chi R} \varphi^*\Big) + \bar{\Psi}_{\ell
  L}i\gamma^{\mu} \Big(\partial_{\mu} +
i\,g\,\frac{1}{2}\,\vec{\tau}\cdot \vec{W}_{\mu} -
i\,g'\,\frac{1}{2}\,B_{\mu} + ie_{\chi} C_{\mu}\Big)\Psi_{\ell L}
\nonumber\\ \hspace{-0.3in}&& - 2 \zeta_e \Big(\bar{\Psi}_{\ell
  L}\psi_{eR}\phi\,\varphi +
\varphi^*\phi^{\dagger}\bar{\psi}_{eR}\Psi_{\ell L}\Big) + 2
\sqrt{2}\,\xi_{\chi}\Big( \varphi^*\bar{\psi}_{n
  R}i\gamma^{\mu}(\partial_{\mu} + i e_{\chi}C_{\mu})\psi_{\chi R} -
i(\partial_{\mu} - i e_{\chi}C_{\mu})\bar{\psi}_{\chi R}\gamma^{\mu}
\psi_{n R} \varphi\Big),
\end{eqnarray}
where $C_{\mu\nu} = \partial_{\mu}C_{\nu} - \partial_{\nu}C_{\mu}$ is
the field strength tensor operator of the dark matter spin--1 gauge
boson field $C_{\mu}$, $\psi_{\chi L} = P_L\psi_{\chi}$ and
$\psi_{\chi R} = P_R \psi_{\chi}$ are the field operators of the left-
and right-handed dark matter fermions $\chi$, $e_{\chi}$ is a gauge
coupling constant or the dark matter {\it charge} of the right--handed
dark matter fermion $\chi$ and the left--handed SM electron and
neutrino. The dark matter {\it hypercharge} $Y'$ is equal to $Y' = +1
$ for the right-handed dark matter fermion $\psi_{\chi R}$, for the
left-handed leptons $\Psi_{\ell L}$ and the complex scalar field
$\varphi$, and $Y' = 0$ for the left-handed dark matter fermion
$\psi_{\chi L}$, respectively.

The parameters $\kappa^2$ and $\gamma$ define a non--vanishing vacuum
expectation value of the dark matter scalar field $\varphi$, that
leads to a non--vanishing mass $m_{\chi}$ of the dark matter fermion
$\chi$, which should be proportional to the coupling constant
$f_{\chi}$. Then, the coupling constant $\xi_{\chi}$ defines a mixing
of the right--handed neutron with the right--handed dark matter
fermion. The term, proportional to the coupling constant $\zeta_e$,
redefines the electron mass in ${\cal L}_{\rm L \sigma M \& SET}$ as
follows
\begin{eqnarray}\label{eq:A.8}
\sqrt{2}\, g_e\Big(\bar{\Psi}_{\ell L}\psi_{eR}\phi +
\phi^{\dagger}\bar{\psi}_{eR}\Psi_{\ell L}\Big) \to 2 \zeta_e
\Big(\bar{\Psi}_{\ell L}\psi_{eR}\phi\,\varphi +
\varphi^*\phi^{\dagger}\bar{\psi}_{eR}\Psi_{\ell L}\Big).
\end{eqnarray}
The Lagrangian Eq.(\ref{eq:A.7}) is invariant under $U'_{Y'}(1)$ dark
matter gauge transformations
\begin{eqnarray}\label{eq:A.9}
&&\psi_{\chi R} \to \psi'_{\chi R} = e^{\,i\alpha_{\chi}}\psi_{\chi
    R}\;\;,\;\;\Psi_{\ell L} \to \Psi'_{\ell L} = e^{\,i\alpha_{\chi}}
  \Psi_{\ell L} \;\;,\;\; \varphi \to \varphi' =
  e^{\,i\alpha_{\chi}}\varphi \;\;,\;\; C_{\mu} \to C'_{\mu} = C_{\mu}
  - \frac{1}{e_{\chi}}\,\partial_{\mu}\alpha_{\chi},
  \nonumber\\ &&\psi_{\chi L} \to \psi'_{\chi L} = \psi_{\chi L}
  \;\;,\;\; \psi_{n R} \to \psi'_{nR} = \psi_{nR} \;\;,\;\;
  \vec{W}_{\mu} \to \vec{W}\,'_{\mu} = \vec{W}_{\mu}\;\;,\;\; B_{\mu}
  \to B'_{\mu} = B_{\mu}.
\end{eqnarray}
Thus, the total Lagrangian of the effective quantum field theory of
strong, electroweak and dark matter interactions we take in the
following form
\begin{eqnarray}\label{eq:A.10}
\hspace{-0.15in} &&{\cal L}_{\rm L \sigma M \& SET \& DM'} =
\bar{\Psi}_{NL}i\gamma^{\mu}\Big(\partial_{\mu} +
i\,g\,\frac{1}{2}\,\vec{\tau}_L \cdot \vec{W}_{\mu} +
i\,g'\,\frac{1}{2}\,B_{\mu} \Big)\Psi_{NL} +
\bar{\psi}_{pR}i\gamma^{\mu}\big(\partial_{\mu} +
i\,g'\,B_{\mu})\psi_{pR} \nonumber\\
\hspace{-0.15in} &&+
\bar{\psi}_{nR}i\gamma^{\mu}\partial_{\mu}\psi_{nR} - \sqrt{2}\,g_{\pi
  N}\big(\bar{\Psi}_{NL}\Phi\,\psi_{pR} +
\bar{\psi}_{pR}\Phi^{\dagger}\Psi_{NL}\big) - \sqrt{2}\, g_{\pi
  N}\big(\bar{\Psi}_{NL}\Phi^c\,\psi_{nR} + \bar{\psi}_{nR}\Phi^{c
  \dagger}\Psi_{NL}\big)\nonumber\\
\hspace{-0.15in} && + \Big(\partial_{\mu}\Phi^{\dagger} -
i\,g\,\frac{1}{2}\,\Phi^{\dagger}\,\vec{\tau}_L\cdot \vec{W}_{\mu} +
i\,g'\,\frac{1}{2}\,\Phi^{\dagger} B_{\mu}
\Big)\Big(\partial^{\mu}\Phi + i\,g\,\frac{1}{2}\,\vec{\tau}_L\cdot
\vec{W}_{\mu}\Phi - i\,g'\,\frac{1}{2}\,B_{\mu}\Phi \Big) + \mu^2\,
\Phi^{\dagger}\Phi - \frac{1}{2}\,\gamma\,
\big(\Phi^{\dagger}\Phi\big)^2\nonumber\\
\hspace{-0.15in} &&  -
\frac{1}{4}\,\vec{W}_{\mu\nu}\cdot \vec{W}^{\mu\nu} -
\frac{1}{4}\,B_{\mu\nu} B^{\mu\nu} + \bar{\psi}_{eR} i \gamma^{\mu}
\big(\partial_{\mu} - i\,g'\,B_{\mu}\big) \psi_{eR} +
\Big(\partial_{\mu}\phi^{\dagger} -
i\,g\,\frac{1}{2}\,\phi^{\dagger}\,\vec{\tau}\cdot \vec{W}_{\mu} -
i\,g'\,\frac{1}{2}\,\phi^{\dagger} B_{\mu}\Big)\nonumber\\
\hspace{-0.15in} && \times \Big(\partial^{\mu}\phi +
i\,g\,\frac{1}{2}\,\vec{\tau}\cdot \vec{W}^{\,\mu}\phi +
i\,g'\,\frac{1}{2}\,B^{\,\mu}\phi\Big) + \tilde{\mu}^2\,
\phi^{\dagger}\phi - \tilde{\lambda}\,(\phi^{\dagger}\phi)^2 +
\bar{\psi}_{\chi R}i\gamma^{\mu}(\partial_{\mu} + i
e_{\chi}C_{\mu})\psi_{\chi R} - \frac{1}{4}\,C_{\mu\nu}C^{\mu\nu}
\nonumber\\
\hspace{-0.15in}&& + (\partial_{\mu} -
ie_{\chi}C_{\mu})\varphi^*(\partial_{\mu} + ie_{\chi}C_{\mu})\varphi +
\kappa^2|\varphi|^2 - \gamma |\varphi|^4 + \bar{\psi}_{\chi
  L}i\gamma^{\mu}\partial_{\mu}\psi_{\chi L} - \sqrt{2}\,
f_{\chi}\big(\bar{\psi}_{\chi R}\psi_{\chi L} \varphi +
\bar{\psi}_{\chi L}\psi_{\chi R} \varphi^*\big) \nonumber\\ && +
\bar{\Psi}_{\ell L}i\gamma^{\mu} \Big(\partial_{\mu} +
i\,g\,\frac{1}{2}\,\vec{\tau}\cdot \vec{W}_{\mu} -
i\,g'\,\frac{1}{2}\,B_{\mu} + ie_{\chi} C_{\mu}\Big)\Psi_{\ell L} - 2
\zeta_e (\bar{\Psi}_{\ell L}\psi_{eR}\phi\,\varphi +
\varphi^*\phi^{\dagger}\bar{\psi}_{eR}\Psi_{\ell L}) \nonumber\\ && +
2 \sqrt{2}\,\xi_{\chi}\Big( \varphi^*\bar{\psi}_{n
  R}i\gamma^{\mu}(\partial_{\mu} + i e_{\chi}C_{\mu})\psi_{\chi R} -
i(\partial_{\mu} - i e_{\chi}C_{\mu})\bar{\psi}_{\chi R}\gamma^{\mu}
\psi_{n R} \varphi\Big).
\end{eqnarray}
This Lagrangian is invariant under transformations of the gauge
$SU_L(2) \times U_Y(1) \times U'_{Y'}(1)$ group \cite{Ivanov2018d}. It
is well-known \cite{Weinberg1967, Weinberg1971} that in the
spontaneously broken or physical phase the Lagrangians like ${\cal
  L}_{\rm L \sigma M \& SET \& DM'}$ should contain only physical
states. It is exactly shown in \cite{Ivanov2019b, Ivanov2020a,
  Ivanov2018d}. The spontaneously broken phase for the
L$\sigma$M$\&$SET part of the Lagrangian ${\cal L}_{\rm L \sigma M \&
  SET \& DM'}$ has been investigated in details in \cite{Ivanov2019b,
  Ivanov2020a}. In turn, for the ${\cal L}_{\rm DM'}$ the
spontaneously broken phase has been analyzed in
\cite{Ivanov2018d}. Here we reproduce the results obtained in
\cite{Ivanov2018d} (see Eqs.(34) - (38) in Ref.\cite{Ivanov2018d}).

\begin{figure}
\centering \includegraphics[height=0.018\textheight]{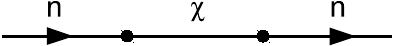}
  \caption{The Feynman diagrams, describing the contribution of the $n
    \longleftrightarrow \chi$ transitions to the neutron mass.}
\label{fig:fig4a}
\end{figure}

In order to define the dark matter sector in the spontaneously broken
or physical phase we take the complex dark matter scalar field
$\varphi$ in the following form $\varphi =
e^{i\,\alpha_{\chi}}\rho/\sqrt{2}$ \cite{Kibble1967,Kibble2015}, where
$\rho$ is a dark matter scalar field, and make a gauge transformation
$\psi_{\chi R} \to e^{i\,\alpha_{\chi}}\psi_{\chi R}$ and $\Psi_{\ell
  L} \to e^{i\,\alpha_{\chi}}\Psi_{\ell L}$
\cite{Kibble1967,Kibble2015}. As a result we arrive at the Lagrangian
\begin{eqnarray}\label{eq:A.11}
\hspace{-0.3in}{\cal L}_{\rm DM'} &=& \bar{\psi}_{\chi
  R}i\gamma^{\mu}(\partial_{\mu} + i \partial_{\mu}\alpha_{\chi} + i
e_{\chi}C_{\mu})\psi_{\chi R} + \bar{\psi}_{\chi
  L}i\gamma^{\mu}\partial_{\mu} \psi_{\chi L} -
f_{\chi}\bar{\psi}_{\chi}\psi_{\chi}\rho -
\frac{1}{4}\,C_{\mu\nu}C^{\mu\nu} + \frac{1}{2}\partial_{\mu}\rho
\partial^{\mu}\rho \nonumber\\ \hspace{-0.3in}&+& \frac{1}{2}(
\partial_{\mu}\alpha_{\chi} + e_{\chi}C_{\mu})(
\partial^{\mu}\alpha_{\chi} + e_{\chi}C^{\mu})\rho^2 +
\frac{1}{2}\,\kappa^2\rho^2 - \frac{1}{4}\,\gamma \rho^4 +
\bar{\Psi}_{\ell L}i\gamma^{\mu} \big(\ldots + i
\partial_{\mu}\alpha_{\chi} + ie_{\chi} C_{\mu}\big)\Psi_{\ell L}
\nonumber\\ \hspace{-0.3in}&+& 2
\sqrt{2}\,\xi_{\chi}\rho\,\Big(\bar{\psi}_{n
  R}i\gamma^{\mu}(\partial_{\mu} + i \partial_{\mu}\alpha_{\chi} + i
e_{\chi}C_{\mu})\psi_{\chi R} - i(\partial_{\mu} - i
\partial_{\mu}\alpha_{\chi} - i e_{\chi}C_{\mu})\bar{\psi}_{\chi
  R}\gamma^{\mu} \psi_{n R}\Big)\nonumber\\ \hspace{-0.3in}&-&
\sqrt{2}\,\zeta_e \rho\, (\bar{\Psi}_{\ell L} \psi_{eR}\phi +
\phi^{\dagger}\bar{\psi}_{eR}\Psi_{\ell L}),
\end{eqnarray}
where the ellipses denotes the contribution of the electroweak gauge
bosons.  The field $C_{\mu} + \partial_{\mu}\alpha_{\chi}/e_{\chi}$
can be treated as a new dark matter spin--1 $Z'$ field
\cite{Kibble1967}. In terms of the $Z'$--field the Lagrangian
Eq.(\ref{eq:A.11}) reads
\begin{eqnarray}\label{eq:A.12}
&&{\cal L}_{\rm DM'} =
  \bar{\psi}_{\chi}i\gamma^{\mu}\partial_{\mu}\psi_{\chi} -
  e_{\chi}\bar{\psi}_{\chi R}\gamma^{\mu}\psi_{\chi R} Z'_{\mu} -
  \frac{1}{4}\,Z'_{\mu\nu}Z'^{\mu\nu} +
  \frac{1}{2}\,e^2_{\chi}Z'_{\mu} Z'^{\mu}\rho^2 -
  f_{\chi}\,\bar{\psi}_{\chi}\psi_{\chi} \rho +
  \frac{1}{2}\partial_{\mu}\rho\partial^{\mu}\rho\nonumber\\ && -
  V(\rho) + 2\,\xi_{\chi} \rho\,\big(\bar{\psi}_{n
    R}i\gamma^{\mu}(\partial_{\mu} + i e_{\chi} Z'_{\mu}) \psi_{\chi
    R} - i(\partial_{\mu} - i e_{\chi} Z'_{\mu})\bar{\psi}_{\chi
    R}\gamma^{\mu} \psi_{n R}\big) - e_{\chi}\bar{\Psi}_{\ell L}
  \gamma^{\mu} \Psi_{\ell L} Z'_{\mu}\nonumber\\ &&- \sqrt{2}\,\zeta_e
  \,\rho\, \Big(\bar{\Psi}_{\ell L} \psi_{eR}\phi +
  \phi^{\dagger}\bar{\psi}_{eR}\Psi_{\ell L}\Big) + \ldots,
\end{eqnarray}
where $Z'_{\mu\nu} = \partial_{\mu}Z'_{\nu} - \partial_{\nu}Z'_{\mu}$
is the field strength tensor operator of the dark matter spin--1 field
$Z'$.  The ellipses denotes the contributions of the interactions of
the electron and neutrino with the electroweak bosons. The potential
of the dark matter scalar $\rho$--field $V(\rho) = -
\frac{1}{2}\,\kappa^2 \rho^2 + \frac{1}{4}\,\gamma \rho^4$ possesses a
minimum at $\langle \rho \rangle = v_{\chi} =
\sqrt{\kappa^2/\gamma}$. Introducing a new scalar field $\rho =
v_{\chi} + S$ \cite{Kibble1967} and taking the SM part of the
effective low-energy system, described by the Lagrangian ${\cal
  L}_{\rm L\sigma M \& SET \& DM'}$, in the spontaneously broken or
physical phase, we transcribe the Lagrangian ${\cal L}_{\rm L\sigma M
  \& SET \& DM'}$ into the form
\begin{eqnarray}\label{eq:A.13}
&&{\cal L}_{\rm L\sigma M \& SET \& DM'} =
  \bar{\psi}_{\chi}\big(i\gamma^{\mu}\partial_{\mu} -
  m_{\chi}\big)\psi_{\chi} + \bar{\psi}_n
  \big(i\gamma^{\mu}\partial_{\mu} - m_n\big) \psi_n + \bar{\psi}_e
  \big(i\gamma^{\mu}\partial_{\mu} - m_e\big) \psi_e +
  \bar{\psi}_{\nu_e} i\gamma^{\mu}\partial_{\mu}\psi_{\nu_e} -
  \frac{1}{4}\,Z'_{\mu\nu} Z'^{\mu\nu} \nonumber\\ &&+
  \frac{1}{2}\,M^2_{Z'}Z'_{\mu} Z'^{\mu} +
  g_{\chi}\Big(\bar{\psi}_ni\gamma^{\mu}(\partial_{\mu} + i e_{\chi}
  Z'_{\mu})(1 + \gamma^5)\psi_{\chi} - i(\partial_{\mu} - i e_{\chi}
  Z'_{\mu})\bar{\psi}_{\chi}\gamma^{\mu}(1 + \gamma^5) \psi_n\Big) -
  e_{\chi}\bar{\psi}_{\chi R}\gamma^{\mu}\psi_{\chi R} Z'_{\mu}
  \nonumber\\ && - e_{\chi} \bar{\Psi}_{eL}\gamma^{\mu}\Psi_{eL}
  Z'_{\mu} + \xi_{\chi}
  S\,\Big(\bar{\psi}_ni\gamma^{\mu}(\partial_{\mu} + i e_{\chi}
  Z'_{\mu})(1 + \gamma^5)\psi_{\chi} - i(\partial_{\mu} - i e_{\chi}
  Z'_{\mu})\bar{\psi}_{\chi}\gamma^{\mu}(1 + \gamma^5) \psi_n\Big) +
  e^2_{\chi} v_{\chi} Z'_{\mu} Z'^{\mu}\, S\nonumber\\ && +
  \frac{1}{2}\,e^2_{\chi} Z'_{\mu} Z'^{\mu}\,S^2 -
  \sqrt{2}\,f_{\chi}\bar{\psi}_{\chi}\psi_{\chi} S +
  \frac{1}{2}\partial_{\mu} S \partial^{\mu} S -
  \frac{1}{2}\,m^2_S\, S^2 - \gamma v_{\chi} S^3 -
  \frac{1}{4}\,\gamma S^4 + \ldots,
\end{eqnarray}
where the ellipses denotes the contributions, which are not important
for the analysis of the neutron- and lepton-dark matter
interactions. Then, $g_{\chi} = \xi_{\chi} v_{\chi}$ and $m_{\chi}$,
$M_{Z'}$ and $m_S$ are masses of the dark matter fermion $\chi$, dark
matter spin--1 $Z'$ and dark matter scalar $S$ fields, respectively,
equal to
\begin{eqnarray}\label{eq:A.14}
m_{\chi} = f_{\chi}\,v_{\chi}\quad,\quad M_{Z'} = e_{\chi} v_{\chi}
\quad,\quad m_S = \sqrt{2 \gamma} v_{\chi}.
\end{eqnarray}
In principle, a mass of the dark matter scalar $S$--field is
arbitrary. In order to allow the $n \to \chi$ transitions only by
virtue of the dark matter spin--1 boson $Z'$ we may delete the
$S$--field from its interactions taking the limit $m_S \to
\infty$. This agrees well with the Appelquist--Carazzone decoupling
theorem \cite{Appelquist1975}. Indeed, keeping the ratio $v_{\chi} =
\sqrt{\kappa^2/\gamma}$ fixed one may set $\gamma \to \infty$. This is
similar to the removal of the scalar $\sigma$--meson from its
interactions in the linear $\sigma$--model of strong low--energy
interactions \cite{Weinberg1967a} (see also \cite{Ivanov2019b,
  Ivanov2020a} and \cite{Ecker1995}).  As a result, the Lagrangian
Eq.(\ref{eq:A.13}) becomes equal to
\begin{eqnarray}\label{eq:A.15}
{\cal L}_{\rm L\sigma M \& SET \& DM'} &=&
\bar{\psi}_{\chi}\big(i\gamma^{\mu}\partial_{\mu} -
m_{\chi}\big)\psi_{\chi} + \bar{\psi}_n
\big(i\gamma^{\mu}\partial_{\mu} - m_n\big) \psi_n + \bar{\psi}_e
\big(i\gamma^{\mu}\partial_{\mu} - m_e\big) \psi_e +
\bar{\psi}_{\nu_e} i\gamma^{\mu}\partial_{\mu}\psi_{\nu_e}
\nonumber\\ &-& \frac{1}{4}\,Z'_{\mu\nu} Z'^{\mu\nu} +
\frac{1}{2}\,M^2_{Z'}Z'_{\mu} Z'^{\mu} + {\cal L}_{\rm n\chi \ell} +
\ldots,
\end{eqnarray}
where ${\cal L}_{\rm n\chi \ell}$ is the Lagrangian of the effective
low-energy neutron-lepton-dark matter interactions given by
\cite{Ivanov2018d}
\begin{eqnarray}\label{eq:A.16}
{\cal L}_{\rm n\chi \ell} &=& g_{\chi}\Big(\bar{\psi}_ni\gamma^{\mu}(1
+ \gamma^5)\partial_{\mu}\psi_{\chi} -
\partial_{\mu}\bar{\psi}_{\chi}i\gamma^{\mu}(1 + \gamma^5)\psi_n\Big)
- g_{\chi}e_{\chi}\Big(\bar{\psi}_n\gamma^{\mu}(1 +
\gamma^5)\psi_{\chi} + \bar{\psi}_{\chi}\gamma^{\mu}(1 +
\gamma^5)\psi_n\Big)Z'_{\mu}\nonumber\\ &-&
\frac{1}{2}\,e_{\chi}\bar{\psi}_{\chi}\gamma^{\mu}(1 + \gamma^5)
\psi_{\chi} Z'_{\mu} - \frac{1}{2}\,e_{\chi}
\bar{\psi}_e\gamma^{\mu}(1 - \gamma^5) \psi_e Z'_{\mu} -
\frac{1}{2}\,e_{\chi} \bar{\psi}_{\nu_e}\gamma^{\mu}(1 - \gamma^5)
\psi_{\nu_e} Z'_{\mu}.
\end{eqnarray}
In \cite{Ivanov2018d} we have estimated the couping constant
$g_{\chi}$ by using the finite contribution to the neutron mass of the
first term in the effective low-energy neutron- and lepton-dark matter
interactions described by the Lagrangian Eq.(\ref{eq:A.16}). The
Feynman diagram of such a contribution is shown in
Fig.\,\ref{fig:fig4a}.
Having assumed that such a contribution should be smaller than the
experimental error $\pm 6 \times 10^{-6}\,{\rm MeV}$ of the neutron
mass \cite{PDG2020}, we have got $|g_{\chi}| < 2.45\times
10^{-3}\,\sqrt{m_n - m_{\chi}}/m_n$, where masses are measured in
MeV. Recall that the coupling constant $g_{\chi}$ is dimensionless.

\begin{figure}
\centering \includegraphics[height=0.075\textheight]{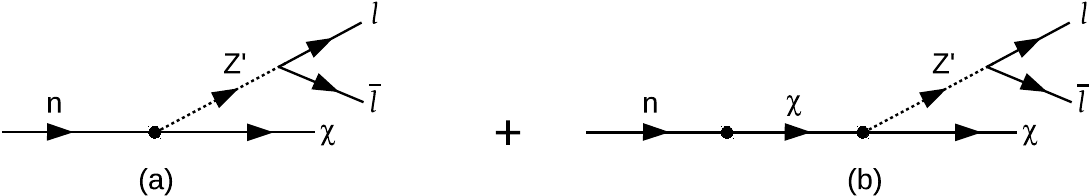}
  \caption{The Feynman diagrams, describing the amplitude of the
    neutron dark matter decays $n \to \chi + \ell + \bar{\ell}$ with
    $\ell = e^-, \nu_e$ and $\bar{\ell} = e^+, \bar{\nu}_e$.}
\label{fig:fig5a}
\end{figure}

The amplitude of the neutron dark matter decays $n \to \chi + \ell +
\bar{\ell}$, where $\ell = e^-, \nu_e$ and $\bar{\ell} = e^+,
\bar{\nu}_e$, is defined by the Feynman diagrams in
Fig.\,\ref{fig:fig5a}.
The analytical expressions for the Feynman diagrams in
Fig.\,\ref{fig:fig5a} are given by
\begin{eqnarray}\label{eq:A.17}
M(n \to \chi + \ell + \bar{\ell})_{\rm Fig.\,\ref{fig:fig5a}a} &=&
\frac{g_{\chi}e^2_{\chi}}{2}[\bar{u}_{\chi}(\vec{k}_{\chi},
  \sigma_{\chi})\gamma^{\mu}(1 + \gamma^5) u_n(\vec{k}_n,
  \sigma_n)]\,\frac{1}{M^2_{Z'} - q^2 - i0}\Big(- \eta_{\mu\nu} +
\frac{q_{\mu}q_{\nu}}{M^2_{Z'}}\Big) \nonumber\\
&&\times\,[\bar{u}_{\ell}(\vec{k}_{\ell})
  \gamma^{\nu}(1 - \gamma^5)v_{\bar{\ell}}(\vec{k}_{\bar{\ell}},
  \sigma_{\bar{\ell}})]
\end{eqnarray}
and 
\begin{eqnarray}\label{eq:A.18}
M(n \to \chi + \ell + \bar{\ell})_{\rm Fig.\,\ref{fig:fig5a}b} &=& -
\frac{g_{\chi}e^2_{\chi}}{2}\,\frac{m^2_n}{m^2_n -
  m^2_{\chi}}\,[\bar{u}_{\chi}(\vec{k}_{\chi},
  \sigma_{\chi})\gamma^{\mu}(1 + \gamma^5) u_n(\vec{k}_n,
  \sigma_n)]\,\frac{1}{M^2_{Z'} - q^2 - i0}\Big( - \eta_{\mu\nu} +
\frac{q_{\mu}q_{\nu}}{M^2_{Z'}}\Big)
\nonumber\\ &&\times\,[\bar{u}_{\ell}(\vec{k}_{\ell}) \gamma^{\nu}(1 -
  \gamma^5)v_{\bar{\ell}}(\vec{k}_{\bar{\ell}}, \sigma_{\bar{\ell}})].
\end{eqnarray}
Summing up the contributions of the Feynman diagrams in
Fig.\,\ref{fig:fig5a} we obtain the amplitude of the neutron dark
matter decays $n \to \chi + \ell + \bar{\ell}$
\begin{eqnarray}\label{eq:A.19}
M(n \to \chi + \ell + \bar{\ell}) &=& - \frac{g_{\chi}e^2_{\chi}}{2
  M^2_{Z'}}\,\frac{m^2_{\chi}}{m^2_n -
  m^2_{\chi}}\,\Big[\bar{u}_{\chi}(\vec{k}_{\chi},
  \sigma_{\chi})\gamma^{\mu}(1 + \gamma^5) u_n(\vec{k}_n,
  \sigma_n)\Big]\,\frac{M^2_{Z'}}{M^2_{Z'} - q^2 - i0}\Big(-
\eta_{\mu\nu} + \frac{q_{\mu}q_{\nu}}{M^2_{\cal
    Z}}\Big)\nonumber\\ &&\times\,[\bar{u}_{\ell}(\vec{k}_{\ell})
  \gamma^{\nu}(1 - \gamma^5)v_{\bar{\ell}}(\vec{k}_{\bar{\ell}},
  \sigma_{\bar{\ell}})].
\end{eqnarray}
Assuming that $M^2_{Z'} \gg q^2$ we arrive at the amplitude
\begin{eqnarray}\label{eq:A.20}
M(n \to \chi + \ell + \bar{\ell}) = \frac{g_{\chi}e^2_{\chi}}{2
  M^2_{Z'}}\,\frac{m^2_{\chi}}{m^2_n -
  m^2_{\chi}}\,\Big[\bar{u}_{\chi}(\vec{k}_{\chi},
  \sigma_{\chi})\gamma^{\mu}(1 + \gamma^5) u_n(\vec{k}_n,
  \sigma_n)\Big]\,[\bar{u}_{\ell}(\vec{k}_{\ell}) \gamma^{\nu}(1 -
  \gamma^5)v_{\bar{\ell}}(\vec{k}_{\bar{\ell}}, \sigma_{\bar{\ell}})],
\end{eqnarray}
which can be obtained from the effective local Lagrangian
\begin{eqnarray}\label{eq:A.21}
&&{\cal L}_{\rm ndm(e + \nu_e)}(x) = -
  \frac{G_F}{\sqrt{2}}\,V_{ud}\,\big[\bar{\psi}_{\chi}(x)\gamma^{\mu}(h_V
    + h_A\gamma^5)\psi_n(x)\big]\,\Big([\psi_e(x)\gamma^{\nu}(1 -
    \gamma^5) \psi_e(x)] + [\psi_{\nu_e}(x)\gamma^{\nu}(1 - \gamma^5)
    \psi_{\nu_e}(x)]\Big),\nonumber\\ &&
\end{eqnarray}
where we have denoted
\begin{eqnarray}\label{eq:A.22}
- \frac{G_F}{\sqrt{2}}V_{ud}h_V = \frac{g_{\chi}e^2_{\chi}}{2
  M^2_{Z'}}\frac{m^2_{\chi}}{m^2_n - m^2_{\chi}}\quad,\quad -
\frac{G_F}{\sqrt{2}}V_{ud} h_A = \frac{g_{\chi}e^2_{\chi}}{2
  M^2_{Z'}}\frac{m^2_{\chi}}{m^2_n - m^2_{\chi}}.
\end{eqnarray}
In terms of the vacuum expectation value of the Higgs--field $v =
1/\sqrt{\sqrt{2}G_F} = 246\,{\rm GeV}$ the coupling constants $h_V$
and $\bar{h}_A$ are defined by
\begin{eqnarray}\label{eq:A.23}
h_V = h_A = - \frac{g_{\chi}}{2
  V_{ud}}\,\frac{v^2}{v^2_{\chi}}\,\frac{m_n}{m_n - m_{\chi}},
\end{eqnarray}
where we have set $m_{\chi} = m_n$ everywhere, except in the mass
difference in the denominator. Taking into account the expression for
$\zeta^{(\rm dm)}$ in Eq.(\ref{eq:9}), $|V_{ud}| = 0.97370$
\cite{PDG2020}, $g_A = 1.2764$ \cite{Abele2018} and $|g_{\chi}| <
2.45\times 10^{-3}\sqrt{m_n - m_{\chi}}/m_n$ we may estimate the
vacuum expectation value $v_{\chi}$. We get $v_{\chi} \sim 0.09\, v
\,(m_n - m_{\chi}) \sim 22\,(m_n - m_{\chi})\,{\rm GeV}$, where $m_n -
m_{\chi}$ is measured in MeV. For such a vacuum expectation value the
mass of the dark matter spin-1 $Z'$-boson is defined by $M_{Z'} \sim
22\,e_{\chi}(m_n - m_{\chi})\,{\rm GeV}$. Setting, for example,
$e_{\chi} = 2$ and assuming that $M_{Z'} \sim 1\,{\rm GeV}$, that is
enough to neglect the contributions of the terms
$q_{\mu}q_{\nu}/M^2_{Z'}$ in Eqs.(\ref{eq:A.17}) and (\ref{eq:A.18}),
we get $(m_n - m_{\chi}) \sim 0.023\,{\rm MeV}$. Apparently, this is
the minimal value of the mass difference $(m_n - m_{\chi})$ in our
approach.

\subsection*{Adler--Bell--Jackiw anomalies and
    violation of renormalizability of renormalizable gauge theories}

We would like to notice that practical applications of the dark matter
sector with $U'_{Y'}(1)$ gauge symmetry to the analysis of different
processes with SM and dark matter particles can be restricted by
tree-- and one--loop approximations. In the one--loop approximation
the dark matter sector with $U'_{Y'}(1)$ gauge symmetry, described by
the Lagrangian Eq.(\ref{eq:A.11}) is renormalizable and gauge
invariant. Renormalizability of the dark matter, described by the
Lagrangian Eq.(\ref{eq:A.11}) is similar to renormalizability of the
heavy baryon chiral perturbation theory (HB$\chi$PT)
\cite{Scherer2011}. The latter also deals with dimensional coupling
constants.

\begin{figure}
\includegraphics[height=0.08\textheight]{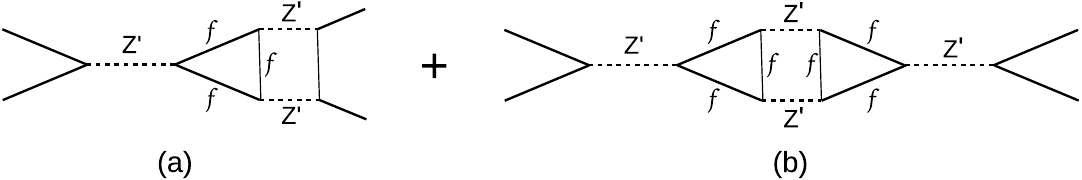}
  \caption{Examples of Feynman diagrams violating renormalizability of
    the dark matter sector with $U'_{Y'}(1)$ gauge symmetry to order
    $O(e^6_{\chi}/2^6)$ (a) and $O(e^8_{\chi}/2^8)$ (b) by the
    Adler--Bell--Jackiw anomaly in processes of fermion--antifermion
    annihilation ($s$-channel) or fermion--fermion scattering
    ($t$-channel).}
\label{fig:fig9}
\end{figure}
\begin{figure}
\includegraphics[height=0.085\textheight]{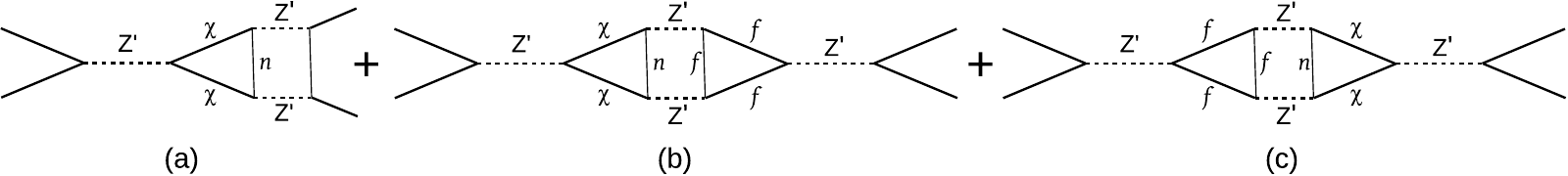}
  \caption{Example of Feynman diagrams violating renormalizability of
    the dark matter sector with $U'_{Y'}(1)$ gauge symmetry to order
    $O(g^2_{\chi}e^4_{\chi}/2^4)$ by the Adler--Bell--Jackiw anomaly,
    caused by the $n\chi Z'$ interaction.}
\label{fig:fig10}
\end{figure}

Renormalizability of the dark matter sector with $U'_{Y'}(1)$ gauge
symmetry can be violated in higher orders of perturbation theory
\cite{Bouchiat1972}--\cite{Bjorken1973} by the Adler--Bell--Jackiw
anomaly \cite{Adler1969,Bell1969}. A lowest order of perturbation
theory, to which renormalizability is violated by the
Adler--Bell--Jackiw anomaly, is $O(e^6_{\chi}/2^6)$. The contributions
of such an order of perturbation theory can appear, for example, in
the processes of fermion--fermion scattering or fermion--antifermion
annihilation. Some examples of Feynman diagrams of order
$O(e^6_{\chi}/2^6)$ and $O(e^8_{\chi}/2^8)$, violating
renormalizability of the amplitudes of fermion--antifermion
annihilation by virtue of the Adler--Bell--Jackiw anomaly in the dark
matter sector with $U'_{Y'}(1)$ gauge symmetry, are shown in Fig.\,
\ref{fig:fig9} (see also Fig.\,14 of Ref. \cite{Bjorken1973}).  The
dark matter spin--1 boson $Z'$ couples to fermions through the vertex
$Z' Z' Z'$ described by one--fermion loops with virtual dark matter
fermions, electrons and neutrinos.  In order to restore
renormalizability to order $O(e^n_{\chi}/2^n)$, where $n \ge 6$, we
propose to add to the Lagrangian Eq.(\ref{eq:A.11}) the term
\begin{eqnarray}\label{eq:A.24}
\delta {\cal L}_{\rm DM'} = \bar{\psi}_{X
  R}i\gamma^{\mu}(\partial_{\mu} + i e_{\chi}C_{\mu})\psi_{X R} -
\sqrt{2}\, f_{X}\big(\bar{\psi}_{X R}\psi_{X L} \varphi +
\bar{\psi}_{X L}\psi_{X R} \varphi^*\big) + \bar{\psi}_{X
  L}i\gamma^{\mu}\partial_{\mu}\psi_{X L},
\end{eqnarray}
where $\psi_{XR} = P_R\psi_X$ and $\psi_{XL} = P_L\psi_X$ are the
field operators of a dark matter fermion $X$. The Lagrangian
Eq.(\ref{eq:A.24}) is invariant under $U'_{Y'}(1)$ dark matter gauge
transformations 
\begin{eqnarray}\label{eq:A.25}
\psi_{X R} \to \psi'_{X R} = e^{\, i \alpha_{\chi}} \psi_{\chi R}\quad,\quad \varphi \to
\varphi' = e^{\,i\alpha_{\chi}} \varphi \quad,\quad \psi_{X L} \to
\psi'_{X L} = \psi_{X L}\quad,\quad C_{\mu} \to
C'_{\mu} = C_{\mu} - \frac{1}{e_{\chi}}\,\partial_{\mu}\alpha_{\chi},
\end{eqnarray}
where $\alpha_{\chi}$ is a gauge parameter. In the spontaneously
broken or physical phase the Lagrangian Eq.(\ref{eq:A.25}) takes the
form
\begin{eqnarray}\label{eq:A.26}
\delta {\cal L}_{\rm DM'} =
\bar{\psi}_X\big(i\gamma^{\mu}\partial_{\mu} - m_X)\psi_X -
\frac{1}{2}\,e_{\chi}\,\bar{\psi}_X\gamma^{\mu}(1 + \gamma^5)\,\psi_X
Z'_{\mu} + \ldots,
\end{eqnarray}
where $m_X = f_X v_{\chi}$ is a mass of the dark matter fermion $X$
such as $m_X = f_Xv_{\chi} \gg m_{\chi}$ and even $m_X \gg
M_{Z'}$. For the dark matter fermion $X$ the dark matter {\it
  hypercharge} $Y'$ is equal to $Y' = + 1$. The anomalous diagrams are
one--loop fermion $Z' Z' Z'$--diagrams of order $e^3_{\chi}/2^3$. The
contributions of dark matter fermions $\chi$ and $X$ give the
Adler--Bell--Jackiw terms with a sign $(-1)$, whereas the electron and
neutrino contributions appear with the sign $(+1)$. Since the
Adler--Bell--Jackiw anomaly does not depend on the mass of virtual
fermions \cite{Adler1969,Bell1969}, the sum of the diagrams with dark
matter fermion $\chi$ and $X$, electron and neutrino loops is free
from the Adler--Bell--Jackiw anomaly.

An additional violation of renormalizability by virtue of the
Adler--Bell--Jackiw anomaly can appear also because of the $n \chi Z'$
interaction.  For example, in the processes of fermion--fermion
scattering and fermion--antifermion annihilation the contribution of
the $n\chi Z'$ interaction, violating renormalizability by virtue of
the Adler--Bell--Jackiw anomaly, is of order $O(e^4_{\chi}
g^2_{\chi}/2^4)$ (see some examples of Feynman diagrams in
Fig.\,\ref{fig:fig10}).  Unfortunately, such a violation of
renormalizability cannot be repaired. It is important to emphasize
that a contribution of the Feynman diagrams, violating
renormalizability by virtue the Adler--Bell--Jackiw anomaly caused by
the $n \chi Z'$ interaction, relative to the main order contribution
$O(e^6_{\chi}/2^6)$ is of order $ O(4g^2_{\chi}/e^2_{\chi}) < 2 \times
10^{-13}$ at $e_{\chi} = 2$ and $g_{\chi} < 2.45 \times
10^{-3}\,\sqrt{m_n - m_{\chi}}/m_n \sim 4 \times 10^{-7}$ for $m_n -
m_{\chi} \simeq 0.023\,{\rm MeV}$. One may argue that violation of
renormalizability to such an order of perturbation theory with
contributions of a relative order $2 \times 10^{-13}$ or even smaller
cannot discredit any quantum field theory model moreover when
practical applications of such a model to the analysis of observable
phenomena can be restricted by the tree- and one-loop approximation
only.

According to \cite{Gross1972}, violation of renormalizability with a
relative order smaller than $2 \times 10^{-13}$ to order
$O(g^2_{\chi}e^4/2^4)$ and higher orders of perturbation theory,
caused by the $n \chi Z'$ interaction, may lead to violation of gauge
invariance only to the same order of magnitude, i.e. to a relative
order smaller than $2 \times 10^{-13}$, and in the same orders of
perturbation theory. So we may argue that up to fourth order of
perturbation theory $O(g^2_{\chi}e^2_{\chi}/2^2) < 2 \times 10^{-13}$
at $e_{\chi} = 2$ (see, for example, the Feynman diagram in
Fig\,\ref{fig:fig10}a without fermion line hooked by two dark matter
spin--1 boson $Z'$), describing fermion--dark matter spin--1 boson
$Z'$ scattering ($f + Z' \to f + Z')$ in the $t$-channel or
fermion--antifermion annihilation into $Z' Z'$--pair ($f + \bar{f} \to
Z' + Z'$) in the $s$-channel and other similar processes,
renormalizability and gauge invariance of the dark matter sector with
$U'_{Y'}(1)$ gauge symmetry are not violated by the $n \chi Z'$
interaction.

\subsection*{Dark matter dynamics in neutron stars}

The influence of the dark matter fermion $\chi$, which can appear in
the final state of the neutron dark matter decays, on dynamics of
neutron stars has been investigated in
\cite{McKeen2018,Baym2018,Motta2018,Cline2018}. The main result is
that dark matter fermions in the equilibrium state with the SM matter
of neutron stars do not destroy the possibility for neutron stars to
reach the maximum mass of about $2\,M_{\odot}$ \cite{Demorest2010},
where $M_{\odot}$ is the mass of the Sun \cite{PDG2020}, only for
$m_{\chi} > 1.2\,{\rm GeV}$. In other words dark matter fermions with
masses $m_{\chi} < m_n$, which can be responsible for the solution of
the neutron lifetime anomaly, are ruled out.

A certain possibility
for existence of dark matter fermions $\chi$ with masses $m_{\chi} <
m_n$ may appear in case of existence of an repulsive interaction
between dark matter fermions mediated by a sufficiently light dark
matter spin--1 bosons, the Compton wavelength of which is larger than
inter-particle distances in neutron stars \cite{McKeen2018}. Such a
possibility for dark matter fermions from the neutron decays has been
realized in scenario by Cline and Cornell \cite{Cline2018} within
$U'(1)$ gauge quantum field theory model with dark matter fermions
$\chi$ coupled to a dark matter photon $A'$, which mass is constrained
by $m_{A'}/g' \leq (45 - 60)\,{\rm MeV}$, where $g'$ is a gauge
coupling constant or a dark matter {\it charge} of dark matter
fermions. According to \cite{Cline2018}, the ratio $m_{A'}/g' \leq (45
- 60)\,{\rm MeV}$ depends on the nuclear equation of state and has
been derived from the requirement for neutron stars to have masses
compatible with $2\,M_{\odot}$ \cite{Demorest2010}.

Since in our approach to the neutron lifetime anomaly the mass of the
dark matter fermion is smaller than the neutron mass $m_{\chi} < m_n$,
we have to accept the mechanism of the influence of dark matter
fermions on dynamics of neutron stars, allowing to have masses of
about $2\,M_{\odot}$, developed by Cline and Cornell
\cite{Cline2018}. For this aim we extend the symmetry of our model
from $SU_L(2) \times U_Y(1) \times U'_{Y'}(1)$ to $SU_L(2) \times
U_Y(1) \times U'_{Y'}(1) \times U''_{Y''}(1)$, where $U''_{Y''}(1)$ is
a new dark matter gauge group. In other words we add to the effective
low-energy field theory, described by the Lagrangian ${\cal L}_{\rm
  L\sigma M \& SET \& DM'}$, the Lagrangian
\begin{eqnarray}\label{eq:A.27}
{\cal L}_{\rm DM''} &=& \bar{\psi}_{\chi
  L}i\gamma^{\mu}(\partial_{\mu} + i
\tilde{e}_{\chi}\tilde{C}_{\mu})\psi_{\chi L} -
\frac{1}{4}\,\tilde{C}_{\mu\nu}\tilde{C}^{\mu\nu} + (\partial_{\mu} -
i \tilde{e}_{\chi}\tilde{C}_{\mu})\tilde{\varphi}^*(\partial_{\mu} + i
\tilde{e}_{\chi}\tilde{C}_{\mu})\tilde{\varphi} + \tilde{\kappa}^2
|\tilde{\varphi}|^2 - \tilde{\gamma}
|\tilde{\varphi}|^4\nonumber\\ &-& 2
\tilde{f}_{\chi}\big(\bar{\psi}_{\chi R}\psi_{\chi L} \varphi
\tilde{\varphi}^* + \bar{\psi}_{\chi L}\psi_{\chi R} \varphi^*
\tilde{\varphi}\big) + \ldots
\end{eqnarray}
invariant under $U''_{Y''}(1)$ gauge transformations, where
$\tilde{C}_{\mu\nu} = \partial_{\mu}\tilde{C}_{\nu} -
\partial_{\nu}\tilde{C}_{\mu}$ is the field strength tensor operator
of the dark matter spin--1 field $\tilde{C}_{\mu}$, $
\tilde{e}_{\chi}$ is the dark matter {\it charge} of the left--handed
dark matter fermions and the dark matter complex scalar field
$\tilde{\varphi}$. The ellipses denotes the contribution of
right-handed dark matter fermions. The Lagrangian Eq.(\ref{eq:A.27})
is invariant under gauge $U''_{Y''}(1)$ transformations. The dark
matter {\it hypercharge} $Y''$ is equal to $Y'' = + 1$ for the
left-handed dark matter fermion field $\psi_{\chi L}$ and the complex
dark matter scalar boson field $\tilde{\varphi}$, and $Y'' = 0$ for
the right-handed dark matter fermion field $\psi_{\chi R}$,
respectively. The first five terms in Eq.(\ref{eq:A.27}) define the
extension of the term $\bar{\psi}_{\chi L}i\gamma^{\mu}\partial_{\mu}
\psi_{\chi L}$ in Eq.(\ref{eq:A.7}). In turn, the last term in
Eq.(\ref{eq:A.27}) is obtained from the term
$f_{\chi}\big(\bar{\psi}_{\chi R}\psi_{\chi L} \varphi +
\bar{\psi}_{\chi L}\psi_{\chi R} \varphi^*\big)$ in Eq.(\ref{eq:A.7})
by the replacement
\begin{eqnarray}\label{e:A.28}
\sqrt{2}\,f_{\chi}\big(\bar{\psi}_{\chi R}\psi_{\chi L} \varphi +
\bar{\psi}_{\chi L}\psi_{\chi R} \varphi^*\big) \to 2
\tilde{f}_{\chi}\big(\bar{\psi}_{\chi R}\psi_{\chi L} \varphi
\tilde{\varphi}^* + \bar{\psi}_{\chi L}\psi_{\chi R} \varphi^*
\tilde{\varphi}\big).
\end{eqnarray}
This implies that the mass of the dark matter fermion $\chi$ appears
in the phase of spontaneously broken $U'_{Y'}(1)\times U''_{Y''}(1)$
symmetry. The Lagrangian Eq.(\ref{eq:A.27}) describes interactions of
dark matter particles only. We would like to notice that the SM
particles and the dark matter particles transforming under
$SU_L(2)\times U_R(1) \times U'_{Y'}(1)$ gauge transformations are
invariant under gauge transformations of the $U''_{Y''}(1)$ group.

Following Kibble \cite{Kibble1967, Kibble2015} and repeating the
procedure expounded above, namely, assuming i) to replace
$\tilde{\varphi}$ by $\tilde{\varphi} =
e^{i\,\tilde{\alpha}_{\chi}}(\tilde{v}_{\chi} +
\tilde{S})/\sqrt{2}$, ii) to make a gauge transformation
$\psi_{\chi L} \to e^{\,i\,\tilde{\alpha}_{\chi}}\psi_{\chi L}$, and
iii) to introduce a new spin--1 boson field $Z''_{\mu} =
\tilde{C}_{\mu} + \partial_{\mu}
\tilde{\alpha}_{\chi}/\tilde{e}_{\chi}$, where $\tilde{v}_{\chi} =
\sqrt{\tilde{\kappa}^2/\tilde{\gamma}}$ is the vacuum expectation
value of the dark matter scalar field $\tilde{\Phi}$, we arrive at the
Lagrangian
\begin{eqnarray}\label{eq:A.29}
{\cal L}_{\rm DM''} &=& \bar{\psi}_{\chi
  L}i\gamma^{\mu}(\partial_{\mu} + i\tilde{e}_{\chi}
Z''_{\mu})\psi_{\chi L} - m_{\chi}\bar{\psi}_{\chi}\psi_{\chi} -
\frac{1}{4}\,Z''_{\mu\nu} Z''^{\mu\nu} + \frac{1}{2}\,M^2_{Z''}
Z''_{\mu} Z''^{\mu} + \tilde{e}^2_{\chi}\tilde{v}_{\chi} Z''_{\mu}
Z''^{\mu}\tilde{S} + \frac{1}{2}\, Z''_{\mu} Z''^{\mu}
\tilde{S}^2\nonumber\\ &+&\frac{1}{2}\,\partial_{\mu}\tilde{S}
\partial^{\mu}\tilde{S} - \frac{1}{2}\,m^2_{\tilde{S}} \tilde{S}^2 +
\tilde{\gamma}\,\tilde{v}_{\chi} \tilde{S}^3 -
\frac{1}{4}\,\tilde{\gamma}\,\tilde{S}^2 -
\tilde{f}_{\chi}\bar{\psi}_{\chi}\psi_{\chi}\Big(v_{\chi}\tilde{S} +
\tilde{v}_{\chi} S + S\tilde{S}\big) + \ldots,
\end{eqnarray}
where $m_{\chi}$, $M_{Z''}$ and $m_{\tilde{S}}$ are masses of the dark
matter fermion $\chi$, dark matter spin--1 $Z''$ and dark matter
scalar $\tilde{S}$ fields
\begin{eqnarray}\label{eq:A.30}
m_{\chi} = \tilde{f}_{\chi}\,v_{\chi}\tilde{v}_{\chi} \quad,\quad
M_{Z''} = \tilde{e}_{\chi} \tilde{v}_{\chi} \quad,\quad m_{\tilde{S}}
= \sqrt{2 \tilde{\gamma}}\, \tilde{v}_{\chi}.
\end{eqnarray}
Without loss of generality we may again set the mass of the dark
matter scalar boson $\tilde{S}$ arbitrary heavy
\cite{Weinberg1967a, Ivanov2019b, Ivanov2020a}. This leads to the
decoupling of the dark matter scalar boson $\tilde{S}$ from the
dark matter fermion $\chi$ and the dark matter spin--1 boson $Z''$ in
agreement with the Appelquist--Carazzone decoupling theorem
\cite{Appelquist1975}.

Since the dark matter spin--1 boson $Z'$ is too heavy to provide a
repulsion at large inter--particle distances in neutron stars, so the
contribution of its repulsion should be taken into account as some
corrections to the repulsion produced by the dark matter spin--1 boson
$Z''$. Indeed, following McKeen {\it et al.}  \cite{McKeen2018} (see
also \cite{Cline2018}), the pressure and energy density of neutron
stars (or the equation of state of neutron stars) should acquire the
corrections (see Eq.(\ref{eq:11}) of Ref. \cite{McKeen2018} and
Eq.(\ref{eq:9}) of Ref.\cite{Cline2018})
\begin{eqnarray}\label{eq:A.31}
\Delta P_{\chi} = \Delta \epsilon_{\chi} =
\frac{1}{2}\,\Big(\frac{\tilde{e}^2_{\chi}}{4 M^2_{Z''}} +
\frac{e^2_{\chi}}{4 M^2_{Z'}}\Big)\,n^2_{\chi} =
\frac{\tilde{e}^2_{\chi}}{8 M^2_{Z''}}\,\Big(1 +
\frac{\tilde{v}^2_{\chi}}{v^2_{\chi}}\Big)\,n^2_{\chi} =
\frac{\tilde{e}^2_{\chi}}{8 M^2_{Z''}}\,\big(1 +
R_{\chi}\big)\,n^2_{\chi}
\end{eqnarray}
caused by the contributions of the dark matter spin--1 bosons $Z''$
and $Z'$, respectively, where we have used $M_{Z''} = \tilde{e}_{\chi}
\tilde{v}_{\chi}$ and $M_{Z'} = e_{\chi} v_{\chi}$. Perturbative
contributions of the dark matter spin--1 boson $Z'$ imply that the
ratio $R_{\chi} = \tilde{v}^2_{\chi}/v^2_{\chi}$ obeys the constraint
$R_{\chi} \ll 1$.  Having neglected the contribution of the dark
matter spin--1 boson $Z'$ to the equation of state we may deal with
the dark matter spin--1 boson $Z''$ only. For the confirmation of such
an approximation we make an estimate of $R_{\chi}$ below
Eq.(\ref{eq:A.34}).

Thus, the part of the total Lagrangian ${\cal L}_{\rm L\sigma M \& SET
  \& DM' \& DM''}$, which should be responsible for dark matter
dynamics in neutron stars, can be written in the following form
\begin{eqnarray}\label{eq:A.32}
{\cal L}_{\rm L\sigma M \& SET \& DM' \& DM''} =
\bar{\psi}_{\chi}(i\gamma_{\mu} - m_{\chi})\psi_{\chi} -
\frac{1}{4}\,Z''_{\mu\nu} Z''^{\mu\nu} + \frac{1}{2}\,M^2_{Z''}
Z''_{\mu} Z''^{\mu} - \frac{1}{2}\,\tilde {e}_{\chi}\,
\bar{\psi}_{\chi}\gamma_{\mu}(1 - \gamma^5)\psi_{\chi} Z''_{\mu} +
\ldots,
\end{eqnarray}
where the ellipsis denotes the contributions of other kinetic and
interaction terms of the SM and dark matter particles, which are not
important for the analysis of the influence of dark matter fermions
with mass $m_{\chi} < m_n$ on the dynamics of neutron stars. In the
non--relativistic approximation the potential of the dark matter
spin--1 boson $Z''$ between two dark matter fermions $\chi$ is equal
to
\begin{eqnarray}\label{eq:A.33}
V_{Z''}(r) = \frac{\tilde{e}^2_{\chi}}{16\pi}\,\frac{\displaystyle
  e^{- M_{Z''}r}}{r}.
\end{eqnarray}
Since it coincides with the potential of the vector field with mass
$M_{Z''}$, describing a repulsive interaction between two fermions
with ``charges'' $\tilde{e}_{\chi}/2$ separated by a distance $r$, we
may apply it for the analysis of dark matter dynamics in neutron stars
in the scenario by Cline and Cornell \cite{Cline2018}. For a short
confirmation of a validity of our model for the analysis of dynamics
of neutron stars we may use the estimate by Cline and Cornell
\cite{Cline2018}. Indeed, according to Cline and Cornell
\cite{Cline2018}, a possibility for neutron stars with dark matter
fermions lighter than neutron and light dark matter spin--1 bosons in
the equilibrium with the SM particles to reach maximum masses
compatible with $2\,M_{\odot}$ places the constraint (see Eq.(12) of
Ref.\cite{Cline2018}). Since the correction to the equation of state
(see Eq.(\ref{eq:A.31})), caused by repulsion between dark matter
fermions with mass $m_{\chi} < m_n$, is fully defined by the dark
matter spin--1 boson $Z''$, the inequality $m_{A'}/g' \leq (45 -
60)\,{\rm MeV}$ (see Eq.(3.4) of Ref.\cite{Cline2018}) should be
saturated only by the dark matter spin--1 boson $Z"$. In our notations
such a constraint reads
\begin{eqnarray}\label{eq:A.34}
\frac{2 M_{Z''}}{\tilde{e}_{\chi}} \lesssim (45 - 60)\,{\rm MeV}.
\end{eqnarray}
This allows to estimate the vacuum expectation value
$\tilde{v}_{\chi}$.  Substituting $M_{Z''} =
\tilde{e}_{\chi}\tilde{v}_{\chi}$ into Eq.(\ref{eq:A.33}) we get
$\tilde{v}_{\chi} \lesssim (23 - 30)\,{\rm MeV}$.  Using
$\tilde{v}_{\chi} \lesssim (23 - 30)\,{\rm MeV}$ and $v_{\chi} \simeq
22\, (m_n - m_{\chi})\,{\rm GeV}$ for the ratio $R_{\chi} =
\tilde{v}^2_{\chi}/v^2_{\chi}$ we get the value $R_{\chi} \sim
1.5\times 10^{-6}/(m_n - m_{\chi})^2$. For $(m_n - m_{\chi}) \sim
0.023\,{\rm MeV}$ we get $R_{\chi} \sim 3 \times 10^{-3}$. Thus, the
contribution of the dark matter spin--1 boson $Z'$ to the equation of
state of neutron stars makes up of about $0.03\,\%$ with respect to
the contribution of the dark matter spin--1 boson $Z''$. This,
confirms our assertion that the contributions of the dark matter
spin--1 boson $Z'$ can be taken into account perturbatively when it is
required.

A specific value of the $Z''$--boson mass depends on the value of the
gauge coupling constant $\tilde{e}_{\chi}$, which can be obtained from
a detailed analysis of the interference of dark matter into dynamics
of neutron stars. Of course, such an analysis, namely i) using the
dark matter fermion mass obeying the constraint $m_{\chi} - m_n \simeq
0.023\,{\rm MeV}$, ii) taking into account the neutron dark matter
decay mode $n \to \chi + \nu_e + \bar{\nu}_e$, where the
neutrino--antineutrino pair possesses a zero net chemical potential
\cite{Baym2018}, and equations of state \cite{Motta2018,Gandolfi2012},
goes beyond the scope of this paper.  We are planning to carry out
this analysis in our forthcoming publications. Here we would like only
to notice that there is practically nothing that can prevent for the
dark matter spin--1 boson $Z''$ to have a mass as light as the dark
matter spin--1 boson $A'$, introduced by Cline and Cornell
\cite{Cline2018}.


\newpage

\begin{thebibliography}{9}
\bibitem{Fornal2018} B. Fornal and B. Grinstein, {\it Dark matter
  interpretation of the neutron decay anomaly}, Phys. Rev. Lett. {\bf
  120}, 191801 (2018); \\ DOI:
  https://doi.org/10.1103/PhysRevLett.120.191801; arXiv:1801.01124
  [hep--ph].

\bibitem{Fornal2019a} B. Fornal and B. Grinstein, {\it Dark side of
  the neutron?}, EPJ Web Conf. {\bf 219}, 05005 (2019); \\ DOI:
  https://doi.org/10.1051/epjconf/201921905005.

 
 \bibitem{Fornal2019b} B. Fornal and B. Grinstein, {\it Dark particle
   interpretation of the neutron decay anomaly},
   J. Phys. Conf. Ser. {\bf 1308}, 012010 (2019); \\ DOI:
   https://doi.org/10.1088/1742-6596/1308/1/012010.

\bibitem{Fornal2020a} B. Fornal and B. Grinstein, {\it Dark matter
  capture by atomic nuclei}, Phys. Lett. B {\bf 811}, 135869 (2020);
  \\ DOI: https://doi.org/10.1016/j.physletb.2020.135869; arXiv:
  2005.04240 [hep-ph].
  
\bibitem{Fornal2020b} B. Fornal and B. Grinstein, {\it Neutron’s dark
  secret}, Mod. Phys. Lett. A {\bf 35}, 2030019 (2020); \\ DOI:
  https://doi.org/10.1142/S0217732320300190; arXiv: 2007.13931
  [hep-ph].

\bibitem{Tang2018} Z. Tang {\it et al.}, {\it Search for the neutron
  decay $n \to X + \gamma$, where $X$ is a dark matter particle},
  Phys. Rev. Lett. {\bf 121}, 022505 (2018); \\ DOI:
  https://doi.org/10.1103/PhysRevLett.121.022505; arXiv:1802.01595 [nucl-ex].

\bibitem{Sun2018} X. Sun {\it et al.} (the UCNA Collaboration), {\it
  Search for dark matter decay of the free neutron from the UCNA
  experiment: $n \to \chi + e^- + e^+$}, Phys. Rev. C {\bf 97}, 052501
  (2018); \\ DOI: https://doi.org/10.1103/PhysRevC.97.052501; arXiv:
  1803.10890 [nucl-ex].

\bibitem{Klopf2019} M. Klopf, E. Jericha, B. M\"arkisch, H. Saul,
  T. Soldner, and H. Abele, {\it Constraints on the dark matter
    interpretation $n \to \chi + e^+ + e^-$ of the neutron decay
    anomaly with the PERKEO II experiment}, Phys. Rev. Lett. {\bf
    122}, 222503 (2019); \\ DOI:
  https://doi.org/10.1103/PhysRevLett.122.222503; arXiv: 1905.01912
  [hep-ex].
  

\bibitem{Ivanov2018d} A. N. Ivanov, R. H\"ollwieser, N. I. Troitskaya,
  M. Wellenzohn, and Ya. A. Berdnikov, {\it Neutron dark matter
    decays}; arXiv: 1806.10107 [hep-ph].

\bibitem{Berezhiani2017a} Z. Berezhiani, {\it Unusual effects in $n-n'$
  conversion}, talk at INT Workshop INT-17-69W, Seattle, 23-27 Oct.
  2017, http://www.int.washington.edu/talks/WorkShops/
  int${_17_6}$9W/People/Berezhiani${_Z}$/Berezhiani3.pdf.
  

\bibitem{Berezhiani2019a} Z. Berezhiani, {\it Neutron lifetime puzzle
  and neutron–mirror neutron oscillation}, Eur. Phys. J. C {\bf 79},
  484 (2019); \\ DOI: https://doi.org/10.1140/epjc/s10052-019-6995-x;
  arXiv: 1807.07906 [hep-ph].

\bibitem{Berezhiani2017b} Z. Berezhiani, M. Frost, Yu. Kamyshkov, B.
  Rybolt, and L. Varriano, {\it Neutron disappearance and regeneration
    from mirror state} Phys. Rev. D {\bf 96}, 035039 (2017); \\ DOI:
  https://doi.org/10.1103/PhysRevD.96.035039; arXiv: 1703.06735
  [hep-ex].

\bibitem{Berezhiani2018} Z. Berezhiani, R. Biondi, P. Geltenbort,
  I. A. Krasnoshchekova, V. E. Varlamov, A. V. Vassiljev, and
  O. M. Zherebtsov, {\it New experimental limits on neutron - mirror
    neutron oscillations in the presence of mirror magnetic field},
  Eur. Phys. J. C {\ bf 78}, 717 (2018); \\ DOI:
  https://doi.org/10.1140/epjc/s10052-018-6189-y; arXiv: 1712.05761
  [hep-ex].
  
\bibitem{Berezhiani2019b} Z. Berezhiani, {\it Neutron lifetime and
  dark decay of the neutron and hydrogen}, LHEP {\bf 2}, 118 (2019);
  \\ DOI: https://doi.org/10.31526/lhep.1.2019.118; arXiv: 1812.11089
     [hep-ph].       
  
\bibitem{PDG2020} P. A. Zyla {\it et al.}, {\it Review of particle
  physics} (Particle Data Group), Prog. Theor. Exp. Phys. {\bf 2020},
  083C01 (2020); \\ DOI: https://doi.org/10.1093/ptep/ptaa104.

\bibitem{DGH2014} J. F. Gonoghue, E. Golowich, and B. R. Holstein, in
  {\it Dynamics of the Standard Model}, 2nd edition, Cambridge
  University Press, Cambridge 2014; \\ DOI:
  https://doi.org/10.1017/CBO9780511803512.  

\bibitem{Ivanov2019b} A. N. Ivanov, R. H\"ollwieser, N. I. Troitskaya,
  M. Wellenzohn, and Ya. A. Berdnikov, {\it Radiative corrections of
    order $O(\alpha E_e/m_N)$ to Sirlin's radiative corrections of
    order $O(\alpha/\pi)$ to the neutron lifetime}, Phys. Rev. D {\bf
    99}, 093006 (2019);\\ DOI:
  https://doi.org/10.1103/PhysRevD.99.093006; arXiv:1905.01178
  [hep-ph].

\bibitem{Ivanov2020a} A. N. Ivanov, R. H\"ollwieser, N. I. Troitskaya,
  M. Wellenzohn, and Ya. A. Berdnikov, {\it Radiative corrections of
    order $O(\alpha E_e/m_N)$ to Sirlin's radiative corrections of
    order $O(\alpha/\pi)$, induced by hadronic structure of the
    neutron}, Phys. Rev. D {\bf 103}, 113007 (2021); \\ DOI:
  https://link.aps.org/doi/10.1103/PhysRevD.103.113007; arXiv:
  2105.06952 [hep-ph].  

\bibitem{McKeen2018} D. McKeen, A. E. Nelson, S. Reddy, and D. Zhou,
  {\it Neutron stars exclude light dark baryons},
  Phys. Rev. Lett. {\bf 121}, 061802 (2018); \\ DOI:
  https://doi.org/10.1103/PhysRevLett.121.061802; arXiv:1802.08244
  [hep-ph].

\bibitem{Baym2018} G. Baym, D. H. Beck, P. Geltenbort, and J. Shelton,
  {\it Testing dark decays of baryons in neutron stars},
  Phys. Rev. Lett. {\bf 121}, 061801 (2028); \\ DOI:
  https://doi.org/10.1103/PhysRevLett.121.061801; arXiv: 1802.08282
  [hep-ph].

\bibitem{Motta2018} T. F. Motta, P. A. M. Guichon, and A. W. Thomas,
  {\it Implications of neutron star properties for the existence of
    light dark matter}, arXiv: 1802.08427 [nucl-th].

\bibitem{Cline2018} J. M. Cline and J. M. Cornell, {\it Dark decay of
  the neutron}, J. High Energ. Phys. {\bf 2018}, 81 (2018); \\ DOI:
  https://doi.org/10.1007/JHEP07(2018)081; arXiv: 1803.04961 [hep-ph].

\bibitem{Demorest2010} P. Demorest, T. Pennucci, S. Ransom, M. Roberts
  and J. Hessels, {\it Shapiro delay measurement of a two solar mass
    neutron star}, Nature {\bf 467}, 1081 (2010); \\ DOI:
  https://doi.org/10.1038/nature09466.


  \bibitem{Abele2018} B. M\"arkisch, H. Mest, H. Saul, X. Wang,
  H. Abele, D. Dubbers, M. Klopf, A. Petoukhov, C. Roick, T. Soldner,
  and D. Werder, {\it Measurement of the weak axial-vector coupling
    constant in the decay of free neutrons using a pulsed cold neutron
    beam}, Phys. Rev. Lett. {\bf 122}, 242501 (2019); \\ DOI:
  https://doi.org/10.1103/PhysRevLett.122.242501; arXiv: 1812.04666
  [nucl-ex].

 \bibitem{Czarnecki2018} A. Czarnecki, W. J. Marciano, and A. Sirlin,
  {\it The neutron lifetime and axial coupling constant connection},
  Phys. Rev. Lett. {\bf 120}, 202002 (2018); \\ DOI:
  https://doi.org/10.1103/PhysRevLett.120.202002; arXiv:1802.01804
  [hep--ph].

 \bibitem{Ivanov2013} A. N. Ivanov, M. Pitschmann, and
  N. I. Troitskaya, {\it Neutron beta decay as a laboratory for
    testing the standard model}, Phys. Rev. D {\bf 88}, 073002 (2013);
  \\ DOI: https://doi.org/10.1103/PhysRevD.88.073002; arXiv:1212.0332
     [hep--ph].
 

\bibitem{Fierz1937} M. Fierz, {\it Zur Fermischen Theorie des
  $\beta$-Zerfalls}, Z. Physik {\bf 104}, 553 (1937); \\ DOI:
  https://doi.org/10.1007/BF01330070.

\bibitem{Jackson1957} J. D. Jackson, S. B. Treiman, and H. W. Wyld
  Jr., {\it Possible tests of time reversal invariance in beta decay},
  Phys. Rev. {\bf 106}, 517 (1957); \\ DOI:
  https://doi.org/10.1103/PhysRev.106.517.

\bibitem{Hardy2020} J. C. Hardy and I. S. Towner, {\it Superallowed
  $0^+ \to 0^+$ nuclear beta decays: 2020 critical survey, with
  implications for $V_{ud}$ and CKM unitarity}, Phys. Rev. C {\bf
  102}, 045501 (2020); \\ DOI:
  https://doi.org/10.1103/PhysRevC.102.045501.

\bibitem{Severijns2019} M. Gonz\'alez--Alonso, O. Naviliat--Cuncic,
  and N. Severijns, {\it New physics searches in nuclear and neutron
    beta decay}, Prog. Part. Nucl. Phys. {\bf 104}, 165 (2019);
  \\ DOI: https://doi.org/10.1016/j.ppnp.2018.08.002.

\bibitem{Abele2019} H. Saul, Ch. Roick, H. Abele, H. Mest, M. Klopf,
  A. Petukhov, T. Soldner, X. Wang, D. Werder, and B.  M\"arkisch,
  {\it Limit on the Fierz interference term b from a measurement of
    the beta asymmetry in neutron decay}, Phys. Rev. Lett. {\bf 125},
  112501 (2020); \\ DOI:
  https://doi.org/10.1103/PhysRevLett.125.112501.

\bibitem{Young2019} V. Cirigliano, A. Garcia, D. Gazit,
  O. Naviliat-Cuncic, G. Savard, and A. Young, {\it Precision beta
    decay as a probe of new physics}, arXiv:1907.02164 [nucl-ex].

\bibitem{Sun2020} X. Sun {\it et al.}, {\it Improved limits on Fierz
  interference using asymmetry measurements from the ultracold neutron
  asymmetry (UCNA) experiment} (UCNA Collaboration), Phys.  Rev. C
  {\bf 101}, 035503 (2020); \\ DOI:
  https://doi.org/10.1103/PhysRevC.101.035503.

\bibitem{Ivanov2019x} A. N. Ivanov, R. H\"ollwieser, N. I. Troitskaya,
  M. Wellenzohn, and Ya. A. Berdnikov, {\it Precision analysis of
    pseudoscalar interactions in neutron beta decays}, Nucl. Phys. B
  {\bf 951}, 114891 (2020); \\ DOI:
  https://doi.org/10.1016/j.nuclphysb.2019.114891; arXiv:1905.04147
  [hep-ph].

\bibitem{Ivanov2019y} A. N. Ivanov, R. H\"ollwieser, N. I. Troitskaya,
  M. Wellenzohn, and Ya. A. Berdnikov, {\it Neutron dark matter decays
    and correlation coefficients of neutron beta decays},
  Nucl. Phys. B {\bf 938}, 114 (2019); \\ DOI:
  https://doi.org/10.1016/j.nuclphysb.2018.11.005.

\bibitem{Gamow1941} G. Gamow and M. Sch\"onberg, {\it Neutrino theory
  of stellar collapse}, Phys. Rev. {\bf 59}, 539 (1941); \\ DOI:
  https://doi.org/10.1103/PhysRev.59.539.

\bibitem{Friman1979} B. L. Friman and O. V. Maxwell, {\it Neutrino
  emissivities of neutron stars}, ApJ {\bf 232}, 541 (1979); \\ DOI:
  10.1086/157313.

\bibitem{Hansel1995} P. H\"ansel, {\it Urca processes in dense matter
 and neutron star cooling"}, Space Science Reviews. {\bf 74}, 427
 (1995); \\ DOI: https://doi.org/10.1007/BF00751429


\bibitem{Ivanov2005} A. N. Ivanov, M. Cargnelli, M. Faber,
  H. Fuhrmann, V. A. Ivanova, J. Marton, N.I. Troitskaya, and
  J. Zmeskal, {\it On kaonic deuterium: Quantum field theoretic and
    relativistic covariant approach}, Eur. Phys. J. A {\bf 23}, 79
  (2005); \\ DOI: https://doi.org/10.1140/epja/i2004-10055-3; arXiv:
  nucl-th/0406053.


\bibitem{Christlmeier2008} S. Christlmeier and H. W. Grie\ss hammer,
  {\it Pion-less effective field theory on low-energy deuteron
    electrodisintegration}, Phys. Rev. C {\bf 77}, 064001 (2008);
  \\ DOI: https://doi.org/10.1103/PhysRevC.77.064001.

\bibitem{Machleidt1987} R. Machleidt, K. Holinde, and C. Elster, {\it
  The Bonn meson exchange model for the nucleon nucleon interaction},
  Phys. Rept. {\bf 149}, 1 (1987); \\ DOI:
  https://doi.org/10.1016/S0370-1573(87)80002-9.

\bibitem{Garson2001} M. Gar{\c c}on and J. W. Van Orden, {\it The
  deuteron: structure and form-factors}, Adv. Nucl. Phys. {\bf 26},
  293 (2001); \\ DOI: https://doi.org/10.1007/0-306-47915-$X{}4$

\bibitem{Ivanov2001} A. N. Ivanov, V. A. Ivanova, H. Oberhummer,
  N. I. Troitskaya, and M. Faber, {\it On the D wave state component
    of the deuteron in the Nambu-Jona-Lasinio model of light nuclei},
  Eur. Phys. J. A {\bf 12}, 87 (2001); \\ DOI:
  https://doi.org/10.1007/s100500170041.


\bibitem{Gilman2002} R. A. Gilman and F. Gross, {\it Electromagnetic
  structure of the deuteron}, J. Phys. G {\bf 28}, R37 (2002); \\ DOI:
  https://doi.org/10.1088/0954-3899/28/4/201.


\bibitem{Leun1982} C. Van der Leun and C. Alderliesten, {\it The
  deuteron binding energy}, Nucl. Phys. A {\bf 380}, 261 (1982);
  \\ DOI: https://doi.org/10.1016/0375-9474(82)90105-1.



\bibitem{Fabian1979} W. Fabian and H. Arenh\"ovel, {\it
  Electrodisintegration of deuteron including nucleon detection in
  coincidence}, Nucl. Phys. A {\bf 314}, 253 (1979); \\ DOI:
  https://doi.org/10.1016/0375-9474(79)90599-2.

\bibitem{Arenhovel1982} H. Arenh\"ovel, {\it On deuteron break-up by
  electrons and the momentum distribution of nucleons on the deuteron},
  Nucl. Phys. A {\bf 384}, 287 (1982); \\ DOI: https://doi.org/10.1016/0375-9474(82)90336-0.

\bibitem{Arenhovel2002} H. Arenh\"ovel, W. Leidemann, and
  E. L. Tomusiak, {\it General multipole expansion of polarization
    observables in deuteron electrodisintegration}, Eur. Phys. J. A
  {\bf 14}, 491 (2002); \\ DOI: 10.1140/epja/i2001-10207-y.

\bibitem{Tamae1987} T. Tamae, H. Kawahara, A. Tanaka, M. Nomura,
  K. Namai, and M. Sugawara, {\it Out-of-plane measurement of the
    $D(e,e'p)$ coincidence cross section}, Phys. Rev. Lett. {\bf 59},
  2919 (1987); \\ DOI: https://doi.org/10.1103/PhysRevLett.59.2919.

\bibitem{Neumann-Cosel2002} P. von Neumann-Cosel, A. Richter,
  G. Schrieder, A. Shevchenko, A. Stiller, and H. Arenh\"ovel, {\it
    Deuteron breakup in the ${^2}{\rm H}(e,e'p)$ reaction at low
    momentum transfer and close to threshold}, Phys. Rev. Lett. {\bf
    88}, 202304 (2002).


\bibitem{HernandezMonteagudo2015} C. Hern\'andez-Monteagudo, Yin-Zhe
  Ma, Fr. S. Kitaura, W.  Wang, R.  G\'enova-Santos, J.
  Mac\'ias-P\'erez, and D. Herranz, {\it Evidence of the missing
    baryons from the kinematic Sunyaev--Zel'dovich effect in Planck
    data}, Phys. Rev. Lett. {\bf 115}, 191301 (2015); \\ DOI:
  https://doi.org/10.1103/PhysRevLett.115.191301.

\bibitem{Ejiri2018} H. Ejiri and J. D. Vergados, {\it Neutron
  disappearance inside the nucleus}, J. Phys. G: Nucl. Part. Phys.,
  {\bf 46}, 025104 (2019); \\ DOI:
  https://doi.org/10.1088/1361-6471/aaf55b; arXiv:1805.04477 [hep-ph].


 	
\bibitem{Karananas2018} G. K. Karananas and A. Kassiteridis, {\it
  Small-scale structure from neutron dark decay}, JCAP, {\bf 09}, 036
  (2018); \\ DOI: 10.1088/1475-7516/2018/09/036; arXiv: 1805.03656
  [hep-ph].

\bibitem{McKeen2021} D. McKeen, M. Pospelov and N. Raj, {\it
  Cosmological and astrophysical probes of dark baryons}, Phys. Rev. D
  {\bf 103}, 115002 (2021); \\ DOI:
  https://doi.org/10.1103/PhysRevD.103.115002; arXiv: 2012.09865
  [hep-ph].

  

\bibitem{GellMann1960} M. Gell-Mann and M. L\'evy, {\it The axial
   vector current in beta decay}, Nuovo Cimento {\bf 16}, 705 (1960);
   \\ DOI: 10.1007/BF02859738.

 \bibitem{Lee1972} H. B. Lee, in {\it Chiral dynamics}, Gordon and
   Breach, New York, 1972.

 \bibitem{Nowak1996} M. Nowak, M. Rho, and I. Zahed, in {\it Chiral
   nuclear dynamics}, World Scientific, Singapore $\bullet$ New Jersey
   $\bullet$ London $\bullet$ Hong Kong, 1996.

 \bibitem{Ivanov2021} A. N. Ivanov, R. H\"ollwieser, N. I. Troitskaya,
   M. Wellenzohn, and Ya. A. Berdnikov, {\it Theoretical description
     of the neutron beta decay in the standard model at the level of
     $10^{-5}$}, Phys. Rev. D {\bf 104}, 033006 (2021);\\ DOI:
   https://doi.org/10.1103/PhysRevD.104.033006; arXiv: 2104.11080
   [hep-ph].
 
   
\bibitem{Weinberg1967} S. Weinberg, {\it A model of leptons},
  Phys. Rev. Lett. {\bf 19}, 1264 (1967); \\ DOI:
  https://doi.org/10.1103/PhysRevLett.19.1264.xs

\bibitem{Weinberg1971} S. Weinberg, {\it Physical processes in
  convergent theory of the weak and electromagnetic interactions},
  Phys. Rev. Lett. {\bf 27}, 1688 (1971); \\ DOI:
  https://doi.org/10.1103/PhysRevLett.27.1688. 
  
  
\bibitem{Weinberg1967a} S. Weinberg, {\it Dynamical approach to
  current algebra}, Phys. Rev. Lett. {\bf 18} 188 (1967); \\ DOI:
  https: //doi.org/10.1103/PhysRevLett.18.188.


\bibitem{Weinberg1968} S. Weinberg, {\it Nonlinear realization of
  chiral symmetry}, Phys. Rev. {\bf 166}, 1568 (1968);\\
  DOI: https://doi.org/10.1103/PhysRev.166.1568.

\bibitem{Weinberg1979} S. Weinberg, {\it Phenomenological
  Lagrangians}, Physica {\bf 96A}, 327 (1979);\\ DOI:
  10.1016/0378-4371(79)90223-1.
  

\bibitem{Gasser1984} J. Gasser and H. Leutwyler, {\it Chiral
  perturbation theory}, Annals of Physics {\bf 158}, 142 (1984);\\ DOI:
   https://doi.org/10.1016/0003-4916(84)90242-2.

\bibitem{Gasser1987} J. Gasser, {\it Chiral perturbation theory and
  effective Lagrangians}, Nucl. Phys. B {\bf 279}, 65 (1987);\\ DOI:
   https://doi.org/10.1016/0550-3213(87)90307-5.

\bibitem{Gasser1988} J. Gasser, M. E. Sainio, and A. S$\check{\rm
  v}$arc, {\it Nucleons in chiral loops}, Nucl. Phys. B {\bf 307}, 779
  (1988);\\ DOI: 10.1016/0550-3213(88)90108-3.

\bibitem{Bernard1992} V. Bernard, N. Kaiser, J. Kambor, and
  Ulf-G. Mei\ss ner, {\it Chiral structure of the nucleon},
  Nucl. Phys. B {\bf 388}, 315 (1992);\\ DOI:
   https://doi.org/10.1016/0550-3213(92)90615-I.

\bibitem{Bernard1995} V. Bernard, N. Kaiser, and Ulf-G. Mei\ss ner,
  {\it Chiral dynamics in nucleons and nuclei}, Int. J. Mod. Phys. E
  {\bf 4}, 193 (1995); \\ DOI:
  https://doi.org/10.1142/S0218301395000092.

\bibitem{Ecker1995} G. Ecker, {\it Chiral perturbation theory },
  Prog. Part. Nucl. Phys. {\bf 35}, 1 (1995);\\ DOI:
   https://doi.org/10.1016/0146-6410(95)00041-G.

\bibitem{Ecker1996} G. Ecker, {\it Low-energy QCD},
  Prog. Part. Nucl. Phys. {\bf 36}, 71 (1996);\\ DOI:
   https://doi.org/10.1016/0146-6410(96)00011-7.


\bibitem{Bijnens1996} J. Bijnens, {\it Chiral Lagrangians and
  Nambu-Jona-Lasinio - like models}, Phys. Rep. {\bf 265}, 369 (1996);\\
  DOI: 10.1016/0370-1573(95)00051-8.

\bibitem{Bernard1997} V. Bernard, N. Kaiser, and Ulf-G. Mei\ss ner,
  {\it Aspects of chiral pion - nucleon physics}, Nucl. Phys. A {\bf
    615}, 483 (1997); \\ DOI:
  https://doi.org/10.1016/S0375-9474(97)00021-3.
  
\bibitem{Fettes1998} N. Fettes, Ulf-G. Mei\ss ner, and S. Steininger,
  {\it Pion-nucleon scattering in chiral perturbation theory (I):
    Isospin-symmetric case}, Nucl. Phys. A {\bf 640}, 199 (1998);
  \\ DOI: https://doi.org/10.1016/S0375-9474(98)00452-7.
  
\bibitem{Gasser2000} J. Gasser, {\it Chiral perturbation theory},
  Nucl. Phys. B (Proc. Suppl.) {\bf 86}, 257 (2000);\\ DOI:
  https://doi.org/10.1016/S0920-5632(00)00573-9
 	  
\bibitem{Scherer2002} S. Scherer, {\it Introduction to chiral
  perturbation theory}, Adv. Nucl. Phys. {\bf 27}, 277 (2003);
  hep-ph/0210398.

\bibitem{Scherer2003} T. Fuchs, J. Gegelia, G. Japaridze, and
  S. Scherer, {\it Renormalization of relativistic baryon chiral
    perturbation theory and power counting}, Phys. Rev. D {\bf 68},
  056005 (2003); \\ DOI:  https://doi.org/10.1103/PhysRevD.68.056005.

\bibitem{Myhrer2005} W. P. Alvarez, K. Kubodera, and F. Myhrer, {\it
  Comparison of the extended linear $\sigma$ model and chiral perturbation
  theory}, Phys. Rev. C {\bf 72}, 038201 (2005);\\ DOI:
  https://doi.org/10.1103/PhysRevC.72.038201.

\bibitem{Bernard2007} V. Bernard and Ulf-G. Mei\ss ner, {\it Chiral
  perturbation theory}, Annu. Rev. Nucl. Part. Sci. {\bf 57}, 33
  (2007); \\ DOI: 10.1146/annurev.nucl.56.080805.140449.

\bibitem{Bernard2008} V. Bernard, {\it Chiral perturbation theory and
  baryon properties}, Prog. Part. Nucl. Phys. {\bf 60}, 82 (2008);\\ 
  DOI:  https://doi.org/10.1016/j.ppnp.2007.07.001.

\bibitem{Scherer2010} S. Scherer, {\it Chiral perturbation theory:
  introduction and recent results in one-nucleon sector},
  Prog. Part. Nucl. Phys. {\bf 61}, 1 (2010);\\  DOI:
   https://doi.org/10.1016/j.ppnp.2009.08.002.

\bibitem{Scherer2011} M. R. Schindler and S. Scherer, {\it Chiral
  effective field theories of the strong interactions},
  Eur. Phys. J. Special Topics {\bf 198}, 95 (2011);\\ DOI:
  https://doi.org/10.1140/epjst/e2011-01485-0.

\bibitem{Bissegger2007} M. Bissegger and A. Fuhrer, {\it A
  renormalization group analysis of leading logarithms in ChPT},
  Eur. Phys. J. C {\bf 51}, 75 (2007): \\ DOI:
  https://doi.org/10.1140/epjc/s10052-007-0292-9.

\bibitem{Sirlin1967} A. Sirlin, {\it General properties of the
  electromagnetic corrections to the beta decay of a physical
  nucleon}, Phys. Rev. {\bf 164}, 1767
  (1967);\\ DOI:https://doi.org/10.1103/PhysRev.164.1767.
  

\bibitem{Sirlin1978} A. Sirlin, {\it Current algebra formulation of
  radiative corrections in gauge theories and the universality of the
  weak interactions}, Rev. Mod. Phys. {\bf 50}, 573 (1978);\\ DOI:
  https://doi.org/10.1103/RevModPhys.50.573.



\bibitem{Shann1971} R. T. Shann, {\it Electromagnetic effects in the
  decay of polarized neutrons}, Nuovo Cimento A {\bf 5}, 591
  (1971).\\ DOI: https://doi.org/10.1007/BF02734566.

  
\bibitem{Ivanov2017} A. N. Ivanov, R. H\"ollwieser, N. I. Troitskaya,
   M. Wellenzohn, and Ya. A. Berdnikov, {\it Precision analysis of
     electron energy spectrum and angular distribution of neutron beta
     decay with polarized neutron and electron}, Phys. Rev. C {\bf
     95}, 055502 (2017); \\ DOI: 10.1103/PhysRevC.95.055502;
   arXiv:1705.07330 [hep-ph].


\bibitem{Ivanov2019a} A. N. Ivanov, R. H\"ollwieser, N. I. Troitskaya,
  M. Wellenzohn, and Ya. A. Berdnikov, {\it Test of the Standard Model
    in neutron beta decay with polarized electrons and unpolarized
    neutrons and protons}, Phys. Rev. D 99, 053004 (2019); \\ DOI:
  10.1103/PhysRevD.99.053004; arXiv:1811.04853 [hep-ph].

 \bibitem{Kibble1967} T. W. B. Kibble, {\it Symmetry breaking in
   non-abelian gauge theories}, Phys. Rev. {\bf 155}, 1554 (1967);
   \\ DOI: https://doi.org/10.1103/PhysRev.155.1554.


\bibitem{Kibble2015} T. W. B. Kibble, {\it History of electroweak
  symmetry breaking}, Journal of Physics: Conf. Series {\bf 626},
  012001 (2015); \\ DOI:
  https://doi.org/10.1088/1742-6596/626/1/012001.
  
\bibitem{Appelquist1975} Th. Appelquist and J. Carazzone, {\it
  Infrared singularities and massive fields}, Phys. Rev. D {\bf 11},
  2856 (1975); \\ DOI: https://doi.org/10.1103/PhysRevD.11.2856.

\bibitem{Bouchiat1972} C. Bouchiat, J. Iliopoulos, and P. Meyer, {\it
  An anomaly free version of Weinberg's model}, Phys. Lett. B {\bf
  38}, 519 (1972); \\ DOI:
  https://doi.org/10.1016/0370-2693(72)90532-1.

\bibitem{Gross1972} D. J. Gross and R. Jackiw, {\it Effect of
  anomalies on quasi-renormalizable theories}, Phys. Rev. D {\bf 6},
  477 (1972); \\ DOI: https://doi.org/10.1103/PhysRevD.6.477.

\bibitem{Bjorken1973} J. D. Bjorken and C. H. Llewellyn Smith, {\it
  Spontaneously broken gauge theories of weak interactions and heavy
  leptons}, Phys. Rev. D {\bf 7}, 887 (1973); \\ DOI:
  https://doi.org/10.1103/PhysRevD.7.887.

\bibitem{Adler1969} S. L.  Adler, {\it Axial vector vertex in spinor
  electrodynamics}, Phys. Rev. {\bf 177}, 2426 (1969); \\ DOI:
  https://doi.org/10.1103/PhysRev.177.2426.

\bibitem{Bell1969} J. S. Bell and R. Jackiw, {\it A PCAC puzzle:
  $\pi^0 \to \gamma\gamma$ in the $\sigma$-model}, Nuovo Cim. A {\bf
  51}, 47 (1969); \\ DOI: https://doi.org/10.1007/BF02823296.
  

\bibitem{Gandolfi2012} S. Gandolfi, J. Carlson, and S. Reddy, {\it
  Maximum mass and radius of neutron stars, and the nuclear symmetry
  energy}, Phys. Rev. C {\bf 85}, 032801 (2012); \\ DOI:
  https://doi.org/10.1103/PhysRevC.85.032801.

  
\end{thebibliography}
\end{document}